\documentclass[final,1p,times]{elsarticle}

\newcommand{\be}{\begin{equation}}
\newcommand{\ee}{\end{equation}}
\newcommand{\ba}{\begin{eqnarray}}
\newcommand{\ea}{\end{eqnarray}}

\usepackage{graphicx,amssymb,amsmath,color}

 \journal{Nuclear Physics A}
\biboptions{sort&compress}
\begin{document}

\begin{frontmatter}

\title{Proton and neutron skins and symmetry energy of mirror nuclei}


\author[mymainaddress]{M.K. Gaidarov\corref{mycorrespondingauthor}}
\address[mymainaddress]{Institute for Nuclear Research and Nuclear Energy,
Bulgarian Academy of Sciences, Sofia 1784, Bulgaria}
\cortext[mycorrespondingauthor]{Corresponding author.}
\ead{gaidarov@inrne.bas.bg}

\author[mysecondaryaddress]{I.~Moumene}
\address[mysecondaryaddress]{High Energy Physics and Astrophysics Laboratory,
Faculty of Science Semlalia, Cadi Ayyad University, P.O.B. 2390,
Marrakesh,Morocco}

\author[mymainaddress]{A.N.~Antonov}

\author[mymainaddress]{D.N.~Kadrev}

\author[mymainaddress1]{P.~Sarriguren}
\address[mymainaddress1]{Instituto de Estructura de la Materia, IEM-CSIC,
Serrano 123, E-28006 Madrid, Spain}

\author[mymainaddress2]{E.~Moya de Guerra}
\address[mymainaddress2]{Departamento de Estructura de la Materia, F\'\i sica
T\'ermica y Electr\'onica, and IPARCOS, Facultad de Ciencias F\'\i
sicas, Universidad Complutense de Madrid, Madrid E-28040, Spain}


\begin{abstract}
The neutron skin of nuclei is an important fundamental property,
but its accurate measurement faces many challenges. Inspired by
charge symmetry of nuclear forces, the neutron skin of a
neutron-rich nucleus is related to the difference between the
charge radii of the corresponding mirror nuclei. We investigate
this relation within the framework of the Hartree-Fock-Bogoliubov
method with Skyrme interactions. Predictions for proton skins are
also made for several mirror pairs in the middle mass range. For
the first time the correlation between the thickness of the
neutron skin and the characteristics related with the density
dependence of the nuclear symmetry energy is investigated
simultaneously for nuclei and their corresponding mirror partners.
As an example, the Ni isotopic chain with mass number $A=48-60$ is
considered. These quantities are calculated within the coherent
density fluctuation model using Brueckner and Skyrme
energy-density functionals for isospin asymmetric nuclear matter
with two Skyrme-type effective interactions, SkM* and SLy4.
Results are also presented for the symmetry energy as a function
of $A$ for a family of mirror pairs from selected chains of nuclei
with $Z=20$, $N=14$, and $N=50$. The evolution curves show a
similar behavior crossing at the $N=Z$ nucleus in each chain and a
smooth growing deviation when $N\neq Z$ starts. Comparison of our
results for the radii and skins with those from the calculations
based on high-precision chiral forces is made.
\end{abstract}


\end{frontmatter}


\section{Introduction}

The neutron-skin thickness represents a ground state property of
finite nucleus that is a strong indicator of the isovector
properties of effective nuclear interactions \cite{Reinhard2010}.
Its knowledge gives more insight into the properties of
neutron-rich nuclei and neutron stars, and the equation of state
(EOS) of asymmetric nuclear matter (ANM). The direct determination
of the neutron-skin thickness usually involves the precise
measurement of the root mean square (rms) radii of both charge and
mass distributions. Electron-nucleus scattering has proven to be
an excellent tool for the study of nuclear structure. In
particular, it has accumulated much reliable information on the
charge density distributions of stable nuclei. Therefore, it is
believed that the new facilities in GSI
\cite{Simon2007,Antonov2011} and RIKEN
\cite{Suda2005,Suda2012,Wakasugi2013} will provide a good
opportunity to study the charge density, and consequently the
proton density distribution, of unstable nuclei by elastic
electron scattering. In RIKEN-SCRIT the first elastic electron
scattering experiment on the stable $^{132}$Xe has already been
performed \cite{Tsukada2017}. Unfortunately, a measurement of the
neutron density distributions to a precision and details
comparable to that of the proton one is hardly possible. It turned
out that to get information on the neutron-skin thickness one
needs data obtained with probes having different sensitivities to
the proton and neutron distributions.

The model-independent measurement of parity-violating asymmetry
\cite{Moreno2009,Donnelly89} (which is sensitive to the neutron
distribution) in the elastic scattering of polarized electrons
from $^{208}$Pb at JLAB within the PREX Collaboration
\cite{prex,Abrahamyan2012} has provided the first electroweak
observation of the neutron-skin thickness $0.33^{+0.16}_{-0.18}$
fm in $^{208}$Pb (see also Ref.~\cite{Gaidarov2012} for more
discussion). A PREX-II experiment has been approved \cite{prexII}
and it is expected to reach the 0.06 fm sensitivity in the neutron
radius of $^{208}$Pb. Parity-violating experiments (CREX) are
planned for the $^{48}$Ca nucleus \cite{CREX,Horowitz2014}.

Experimental difficulties in direct neutron-skin measurements and
uncertainty about the sensitivity of mean-field models to
isovector quantities \cite{Furnstahl2002} make alternative
approaches desirable. Mahzoon {\it et al.} have proposed a method
of determining the neutron rms radius and the neutron-skin
thickness of $^{48}$Ca using a dispersive-optical-model (DOM)
analysis of bound and scattering data to constrain the nucleon
self-energy \cite{Mahzoon2017}. A best fit neutron skin of
$0.249\pm 0.023$ fm was deduced. Previously, applying the DOM the
same authors considered the $N=Z$ system $^{40}$Ca and extracted a
very small, but negative, skin thickness of -0.06 fm
\cite{Mahzoon2014}. It has been argued that while the proton and
neutron distributions of $^{40}$Ca are very similar and there is
essentially no neutron skin, as expected, the magnitude of the
sizable neutron skin of $^{48}$Ca comes predominately from the
$f_{7/2}$ orbital, reflecting its centrifugal barrier. The
nonlocal DOM dispersive analysis of $^{208}$Pb has been also
carried out, from which a neutron skin of $0.25\pm 0.05$ fm was
deduced \cite{Atkinson2020}. A very recent systematic study of
more nuclei has been performed in Ref.~\cite{Pruitt2020} using a
newly-generalized version of the DOM. In addition, the neutron
density distributions and the neutron-skin thickness in $^{96}$Zr
and $^{96}$Ru nuclei have been probed with ultrarelativistic
isobaric collisions \cite{Li2019} showing that they could be
determined to a precision which may exceed those achieved by
traditional low energy nuclear experiments.

It has been shown in Ref.~\cite{Brown2017} that in the case of a
perfect charge symmetry the neutron skin in a given nucleus can be
obtained from the proton radii of mirror nuclei. Therefore,
besides the planned JLAB experiment, measurements of mirror charge
radii could be an alternative with a competitive precision. The
necessary step after measuring the charge rms radii is to apply
the relativistic and finite size corrections to deduce the
point-proton rms radii. The correlations discussed in
Ref.~\cite{Brown2017} between the neutron skin and the difference
of the proton radii are determined for a particular mirror pair.
This was realized by constructing 48 Skyrme functionals to predict
different skins of $^{208}$Pb within a chosen range. Moreover, it
was also shown that the difference in the charge radii of mirror
nuclei is proportional to the slope of the symmetry energy $L$ at
saturation density, even in the presence of the Coulomb
corrections. The same findings have been confirmed in an approach
based on a set of 14 relativistic energy density functionals
(EDFs) spanning a wide region of values of $L$ \cite{Yang2018}. In
a recent work \cite{Sammarruca2018} Sammarruca has applied an
isospin-asymmetric EOS derived microscopically from high-precision
chiral few-nucleon interactions to study these correlations for a
family of mirror pairs.

The nuclear symmetry energy, which is defined as the difference
between the energies of pure neutron and symmetric nuclear matter,
is an important physical quantity in nuclear physics and
astrophysics (see, e.g., \cite{NSE2014,Lattimer2007,Li2008}). It
can account for many experimental facts at low nuclear densities,
especially the existence of neutron skin. The size of the neutron
skin is determined by the relative strengths of the symmetry
energy between the central near-saturation and peripheral
less-dense regions. Therefore the neutron-skin thickness is a
measure of the density dependence of the symmetry energy around
saturation
\cite{Typel2001,Furnstahl2002,Steiner2005,RocaMaza2011}. We have
investigated possible relationships between the neutron-skin
thickness of spherical \cite{Gaidarov2011} and deformed
\cite{Gaidarov2012} neutron-rich nuclei and the symmetry energy
characteristics of nuclear matter for these nuclei. In
Refs.~\cite{Gaidarov2012,Gaidarov2011} the analysis of the nuclear
symmetry energy, the neutron pressure, and the asymmetric
compressibility has been carried out on the basis of the Brueckner
EDF for infinite nuclear matter. The capability of the coherent
density fluctuation model (CDFM) \cite{Ant80,AHP} to provide a
transparent and analytic way for the transition from nuclear
matter to finite nuclei has been demonstrated in these studies.

In the present paper we aim to investigate the relations between
the quantities mentioned above among isotopic and isotonic chains
with different masses. We focus on nuclei in the mass region
$A=48-60$, in which the Ni isotopes and respective mirror nuclei
are studied. As an important task, we search for possible
correlations between the neutron-skin thickness and the EOS
parameters (symmetry energy, pressure, asymmetric compressibility)
for various Ni isotopes and their corresponding mirror partners.
For the case of these isotopes such correlations have already
being investigated in
Refs.~\cite{Gaidarov2011,Gaidarov2012,Gaidarov2014}. Here we also
calculate the proton skins of Argon isotopes ($A=32-40$) and
predictions for them are made, in comparison with the empirical
data \cite{Ozawa2002} and the microscopic results of Sammarruca
\cite{Sammarruca2018}. Also, we inspect the relation between the
neutron skin and the difference of the proton radii of the
corresponding mirror nuclei for the $Z=20$ and $Z=28$ isotopic
chains and for $N=14$ and $N=50$ isotonic chains. Finally, we pay
particular attention to the $Z=20$ isotopic chain, including
$^{48}$Ca, inspired by the new experiment on this nucleus (CREX)
that is ongoing at JLAB \cite{CREX}.

The nuclear densities and radii are calculated within a
self-consistent Hartree-Fock-Bogoliubov (HFB) method by using the
cylindrical transformed deformed harmonic-oscillator basis
(HFBTHO) \cite{Stoitsov2013,Stoitsov2005} that has been adopted
previously in Refs.~\cite{Antonov2017,Antonov2018}. The results
for the symmetry energy and related quantities  in the specified
nuclei are obtained in the CDFM framework by use of Brueckner and
Skyrme EDFs for infinite nuclear matter with two Skyrme-type
effective interactions: SLy4 and SkM*.

The structure of this paper is the following. In
Section~\ref{s:mirror} we present the relation between the neutron
skin of a nucleus and the difference between the proton radii of
the corresponding mirror nuclei in the presence of perfect charge
symmetry. Section~\ref{s:theory} contains the definitions of the
key EOS parameters in nuclear matter and CDFM formalism that
provides a way to calculate the intrinsic quantities in finite
nuclei. The numerical results and discussions are presented in
Section~\ref{s:results}. The summary of the results and main
conclusions of the study are given in Section~\ref{s:conclusions}.

\section{Skins of mirror nuclei \label{s:mirror}}

Mirror nuclei, with interchanged numbers of protons and neutrons,
are expected to have similar nuclear structure due to the isospin
symmetry of nuclear forces. For instance, the level schemes of
mirror nuclei should be identical if isospin symmetry is fully
exact. Of course, some differences appear due to the fact that the
Coulomb interaction breaks the isospin symmetry.

Usually the neutron-skin thickness is associated with the
difference between the neutron and proton rms radii (we note that
in our previous study \cite{Sarriguren2007} the formation of a
neutron skin in even-even isotopes of Ni, Kr, and Sn was analyzed
in terms of various definitions). The neutron- and proton-skin
thicknesses are, correspondingly:
\begin{equation}
\Delta R_{n}=R_{n}(Z,N)-R_{p}(Z,N)
\label{eq:1}
\end{equation}
and
\begin{equation}
\Delta R_{p}=R_{p}(Z,N)-R_{n}(Z,N),
\label{eq:2}
\end{equation}
where $R_{n}(Z,N)$ and $R_{p}(Z,N)$ are the rms radii of the
neutron and proton density distributions with $Z$ protons and $N$
neutrons. As can be seen
\begin{equation}
\Delta R_{n}=-\Delta R_{p}.
\label{eq:3}
\end{equation}
Under the assumption of exact charge symmetry, the neutron radius
of a given nucleus is identical to the proton radius of its mirror
nucleus:
\begin{equation}
R_{n}(Z,N)=R_{p}(N,Z).
\label{eq:4}
\end{equation}

Let us introduce the following difference of the proton radii of a
given nucleus and its mirror one:
\begin{equation}
\Delta R_{mirr}=R_{p}(N,Z)-R_{p}(Z,N).
\label{eq:5}
\end{equation}
Thus, in the case of the exact charge symmetry, using
Eq.~(\ref{eq:4}) in Eq.~(\ref{eq:1}), one can obtain the equality
of $\Delta R_{n}$ and $\Delta R_{mirr}$:
\begin{equation}
\Delta R_{n}=R_{p}(N,Z)-R_{p}(Z,N)=\Delta R_{mirr}.
\label{eq:6}
\end{equation}
Hence, from accurate measure of the charge radii of the mirror
pair nuclei, $(Z,N)$ and $(N,Z)$, the neutron skin thickness of
the $(Z,N)$ nucleus can be obtained, provided Coulomb effects are
properly taken into account in the data analyses. This could be an
alternative to the existing methods to determine neutron and
proton skins.

\section{Nuclear EOS parameters in nuclear matter and finite nuclei \label{s:theory}}

The nuclear matter EOS is conventionally defined as the binding
energy per nucleon and can be approximately expressed as
\begin{equation}
E(\rho,\delta)=E(\rho,0)+S^{ANM}(\rho)\delta^2+O(\delta^4)+
\cdot\cdot\cdot \;\;
\label{eq:07}
\end{equation}
in terms of the isospin asymmetry
$\delta=(\rho_{n}-\rho_{p})/\rho$ ($\rho$, $\rho_{n}$ and
$\rho_{p}$ being the baryon, neutron and proton densities,
respectively) (see, e.g.,
\cite{Gaidarov2011,Gaidarov2012,Diep2003,Chen2011}). In
Eq.~(\ref{eq:07}) $E(\rho,0)$ is the energy of isospin-symmetric
matter. Near the saturation density $\rho_{0}$ the symmetry energy
for ANM, $S^{ANM}(\rho)$, can be expanded as
%
%
%
\begin{equation}
S^{ANM}(\rho)=\frac{1}{2}\left.
\frac{\partial^{2}E(\rho,\delta)}{\partial\delta^{2}} \right
|_{\delta=0}=\left.
a_{4}+\frac{p_{0}^{ANM}}{\rho_{0}^{2}}(\rho-\rho_{0})+\frac{\Delta
K^{ANM}}{18\rho_{0}^{2}}(\rho-\rho_{0})^{2}+ \right.
\cdot\cdot\cdot \;.
\label{eq:7}
\end{equation}
The parameter $a_{4}$ is the symmetry energy at equilibrium
($\rho=\rho_{0}$). The pressure $p_{0}^{ANM}$
\begin{equation}
p_{0}^{ANM}=\rho_{0}^{2}\left.
\frac{\partial{S^{ANM}}}{\partial{\rho}} \right |_{\rho=\rho_{0}}
\label{eq:8}
\end{equation}
and the curvature $\Delta K^{ANM}$
\begin{equation}
\Delta K^{ANM}=9\rho_{0}^{2}\left.
\frac{\partial^{2}S^{ANM}}{\partial\rho^{2}} \right
|_{\rho=\rho_{0}}
\label{eq:9}
\end{equation}
of the nuclear symmetry energy at $\rho_{0}$ govern its density
dependence and thus provide important information on the
properties of the nuclear symmetry energy at both high and low
densities. The widely used "slope" parameter $L^{ANM}$ is related
to the pressure $p_{0}^{ANM}$ [Eq.~(\ref{eq:8})] by
\begin{equation}
L^{ANM}=\frac{3p_{0}^{ANM}}{\rho_{0}}.
\label{eq:10}
\end{equation}

In our previous works
\cite{Gaidarov2011,Gaidarov2012,Antonov2016,Danchev2020}, as well
as in the present paper, the transition from the properties of
nuclear matter to those of finite nuclei has been made using the
CDFM. The model is a natural extension of the Fermi-gas one. It is
based on the $\delta$-function approximation of the generator
coordinate method \cite{Grif57} and includes nucleon-nucleon
correlations of collective type. In the present work CDFM is
applied to our studies of the symmetry energy and related
quantities. In the CDFM the one-body density matrix
$\rho(\mathbf{r},\mathbf{r}^{\prime})$ is a coherent superposition
of the one-body density matrices $\rho^{NM}_{x}({\bf r},{\bf
r^{\prime}})$ for spherical ``pieces'' of nuclear matter
(``fluctons'') with densities $\rho_{x}({\bf
r})=\rho_{0}(x)\Theta(x-|{\bf r}|)$ and $\rho_{0}(x)=3A/4\pi
x^{3}$. It has the form:
\begin{equation}
\rho({\bf r},{\bf r^{\prime}})=\int_{0}^{\infty}dx |{\cal
F}(x)|^{2} \rho^{NM}_{x}({\bf r},{\bf r^{\prime}})
\label{eq:11}
\end{equation}
with
\begin{equation}
\rho^{NM}_{x}({\bf r},{\bf r^{\prime}})=3\rho_{0}(x)
\frac{j_{1}(k_{F}(x) |{\bf r}-{\bf r^{\prime}}|)}{(k_{F}(x)|{\bf
r}-{\bf r^{\prime}}|)} \Theta \left (x-\frac{|{\bf r}+{\bf
r^{\prime}}|}{2}\right ).
\label{eq:12}
\end{equation}
In (\ref{eq:12}) $j_{1}$ is the first-order spherical Bessel
function and
\begin{equation}
k_{F}(x)=\left(\frac{3\pi^{2}}{2}\rho_{0}(x)\right )^{1/3}\equiv
\frac{\beta}{x}
\label{eq:13}
\end{equation}
with
\begin{equation}
\beta=\left(\frac{9\pi A}{8}\right )^{1/3}\simeq 1.52A^{1/3}
\label{eq:14}
\end{equation}
is the Fermi momentum of the nucleons in the flucton with a radius
$x$. It follows from Eqs.~(\ref{eq:11}) and (\ref{eq:12}) that the
density distribution in the CDFM has the form:
\begin{equation}
\rho({\bf r})=\int_{0}^{\infty}dx|{\cal
F}(x)|^{2}\rho_{0}(x)\Theta(x-|{\bf r}|)
\label{eq:15}
\end{equation}
and from (\ref{eq:15}) that in the case of monotonically
decreasing local density ($d\rho/dr\leq 0$) the weight function
$|{\cal F}(x)|^{2}$ can be obtained from a known density
(theoretically or experimentally obtained):
\begin{equation}
|{\cal F}(x)|^{2}=-\frac{1}{\rho_{0}(x)} \left.
\frac{d\rho(r)}{dr}\right |_{r=x}
\label{eq:16}
\end{equation}
with normalization
\begin{equation}
\int_{0}^{\infty}dx |{\cal F}(x)|^{2}=1.
\label{eq:17}
\end{equation}

Following the CDFM scheme, the symmetry energy, the slope, and the
curvature for finite nuclei can be defined weighting these
quantities for nuclear matter by means of the weight function
$|{\cal F}(x)|^{2}$ [Eq.~(\ref{eq:16})]. They have the following
forms (see, for instance, Ref.~\cite{Gaidarov2011}):
\begin{equation}
S(A)=\int_{0}^{\infty}dx|{\cal F}(x)|^{2}S^{ANM}(x),
\label{eq:18}
\end{equation}
\begin{equation}
p_{0}(A)=\int_{0}^{\infty}dx|{\cal F}(x)|^{2}p_{0}^{ANM}(x),
\label{eq:19}
\end{equation}
\begin{equation}
\Delta K(A)=\int_{0}^{\infty}dx|{\cal F}(x)|^{2}\Delta K^{ANM}(x).
\label{eq:20}
\end{equation}
Analytical expressions for the nuclear matter quantities
$S^{ANM}(x)$, $p_{0}^{ANM}(x)$, and $\Delta K^{ANM}(x)$
[Eqs.~(\ref{eq:18})-(\ref{eq:20})] derived on the basis of
Brueckner EDF can be found in Ref.~\cite{Gaidarov2011}. As it was
mentioned before, results for the isotopic and isotonic evolution
of the symmetry energy obtained also by Skyrme EDF will be
presented.

As far as the densities and the rms radii of the mirror nuclei are
concerned, they have been obtained within the Skyrme HFB method.
The HFBTHO code \cite{Stoitsov2013,Stoitsov2005} solves the
nuclear Skyrme HFB problem by using the cylindrical transformed
deformed harmonic-oscillator basis. In our case we perform
spherical calculations taking the cylindrical basis in its
spherical limit. In HFBTHO, the direct term of the Coulomb
potential to the total energy is taken into account.

\section{Results of calculations and discussion \label{s:results}}

\subsection{Predictions for nuclear skins. Results of the relation between
the neutron skin of a nucleus and the difference between the
proton radii of the mirror pair}

We show first in Table~\ref{tab1} the results for the rms radii
and proton skins predictions for $Z=10$ and $Z=18$ isotopic chains
including neutron-deficient even-even isotopes in each chain. The
values of these characteristics are given for the case of the SLy4
effective force. It can be seen that for the neutron-deficient Ne
and Ar nuclei the neutron rms radius is much smaller than the
corresponding charge radius. This has the obvious implication that
the neutron-deficient nuclei possess an extended proton skin,
which is clearly demonstrated in Table~\ref{tab1}. A good
agreement between our results and the relativistic mean-field
(RMF) calculations \cite{Lalazissis98} is achieved for $R_{n}$ and
$R_{c}$, as well as with the RMF predictions for $\Delta
R_{n}=-\Delta R_{p}$ [Eq.~(\ref{eq:3})] in \cite{Geng2004} and the
relativistic Hartree-Bogoliubov model results
\cite{Lalazissis2004}. There is also a good agreement of the
proton skin values derived from the HFBTHO code with the proton
skins obtained in Ref.~\cite{Sammarruca2018} on the base of EOS
with high-precision chiral forces. In particular, the obtained
proton skins of the less neutron-deficient Ne and Ar isotopes are
in the interval covering the estimated theoretical errors in
Ref.~\cite{Sammarruca2018}. The latter include uncertainties due
to the variations of the cutoff parameter and the chosen many-body
method to calculate the skins.

\begin{table}
\caption{Neutron ($R_{n}$), proton ($R_{p}$), matter ($R_{m}$),
and charge ($R_{c}$) rms radii (in fm) calculated with SLy4 force
for $Z=10$ and $Z=18$ isotopic chains. The proton skins $\Delta
R_{p}$ [Eq.~(\ref{eq:2})] (in fm) are also shown in comparison
with the results obtained in Ref.~\cite{Sammarruca2018}.}
\label{tab1}
\begin{center}
\begin{tabular}{cccccccccc}
\hline \noalign{\smallskip}
Nucleus & $Z$ & $N$ & & $R_{n}$ & $R_{p}$ & $R_{m}$ & $R_{c}$ & $\Delta R_{p}$ & $\Delta R_{p}$ \cite{Sammarruca2018} \\
\noalign{\smallskip}\hline\noalign{\smallskip}
$^{16}$Ne       & 10 & 6  & & 2.51 & 2.89 & 2.76 & 3.00 & 0.378 & $0.422\pm 0.022$ \\
$^{18}$Ne       &    & 8  & & 2.67 & 2.85 & 2.77 & 2.96 & 0.175 & $0.186\pm 0.012$ \\
$^{20}$Ne       &    & 10 & & 2.81 & 2.84 & 2.82 & 2.95 & 0.029 & $0.032\pm 0.006$ \\
\noalign{\smallskip}\hline\noalign{\smallskip}
$^{30}$Ar       & 18 & 12 & & 3.00 & 3.32 & 3.20 & 3.42 & 0.323 & $0.352\pm 0.019$ \\
$^{32}$Ar       &    & 14 & & 3.07 & 3.28 & 3.19 & 3.38 & 0.216 & $0.225\pm 0.013$ \\
$^{34}$Ar       &    & 16 & & 3.17 & 3.29 & 3.23 & 3.39 & 0.123 & $0.127\pm 0.012$ \\
$^{36}$Ar       &    & 18 & & 3.26 & 3.31 & 3.29 & 3.40 & 0.046 & $0.046\pm 0.007$ \\
\noalign{\smallskip}\hline
\end{tabular}
\end{center}
\end{table}

In Table~\ref{tab2} we list the values of the proton skins
obtained with SkM* force for the same isotopes considered in
Table~\ref{tab1}. Also shown in Table~\ref{tab2} are the
neutron-skin thickness of the corresponding mirror partners, as
well as the proton radii difference of the mirror pair $\Delta
R_{mirr}$. Indeed, the neutron skins of the corresponding
neutron-rich mirror nuclei are smaller than the proton skins for
comparable values of proton-neutron asymmetry. For a given proton
excess $(Z - N)$ the corresponding proton-skin thickness is
observed to be larger than the neutron-skin thickness for the same
value of neutron excess $(N - Z)$. This is obviously due to the
Coulomb repulsion of protons. We note that the obtained $\Delta
R_{n}$ value of $^{16}$Ne is comparable with the value of
$0.333\pm 0.016$ from Ref.~\cite{Sammarruca2018}. It can be seen
from Tables~\ref{tab1} and \ref{tab2} that a small difference is
observed between the values of the predicted proton skins $\Delta
R_{p}$ for the considered $Z=10$ and $Z=18$ nuclei in respect to
the Skyrme force used in the calculations. In particular, the
values of $\Delta R_{p}$ obtained with SLy4 force are larger than
those with SkM* force in the case of Ne isotopes, while they are
smaller in the case of Ar isotopes.

\begin{table}
\caption{Predicted proton skins $\Delta R_{p}$ [Eq.~(\ref{eq:2})]
(in fm) for the same $Z=10$ and $Z=18$ nuclei presented in
Table~\ref{tab1} (columns 1 and 2), neutron skins $\Delta R_{n}$
[Eq.~(\ref{eq:1})] (in fm) of the corresponding mirror nuclei
(columns 3 and 4), and $\Delta R_{mirr}$ [Eq.~(\ref{eq:6})] (in
fm) (column 5) calculated with SkM* force.}
\label{tab2}
\begin{center}
\begin{tabular}{ccccccccc}
\hline \noalign{\smallskip}
Nucleus & & $\Delta R_{p}$ & & Mirror Nucleus & & $\Delta R_{n}$ & & $\Delta R_{mirr}$ \\
\noalign{\smallskip}\hline\noalign{\smallskip}
$^{16}$Ne      & & 0.366  & & $^{16}$C   & &  0.308 & & -0.360 \\
$^{18}$Ne      & & 0.160  & & $^{18}$O   & &  0.111 & & -0.147 \\
$^{20}$Ne      & & 0.024  & & $^{20}$Ne  & & -0.024 & &  0.000 \\
\noalign{\smallskip}\hline\noalign{\smallskip}
$^{30}$Ar      & & 0.327  & & $^{30}$Mg  & &  0.238 & & -0.308 \\
$^{32}$Ar      & & 0.222  & & $^{32}$Si  & &  0.136 & & -0.188 \\
$^{34}$Ar      & & 0.128  & & $^{34}$S   & &  0.041 & & -0.088 \\
$^{36}$Ar      & & 0.044  & & $^{36}$Ar  & & -0.044 & &  0.000 \\
\noalign{\smallskip}\hline
\end{tabular}
\end{center}
\end{table}

\begin{figure}
\centering
\includegraphics[width=0.6\linewidth]{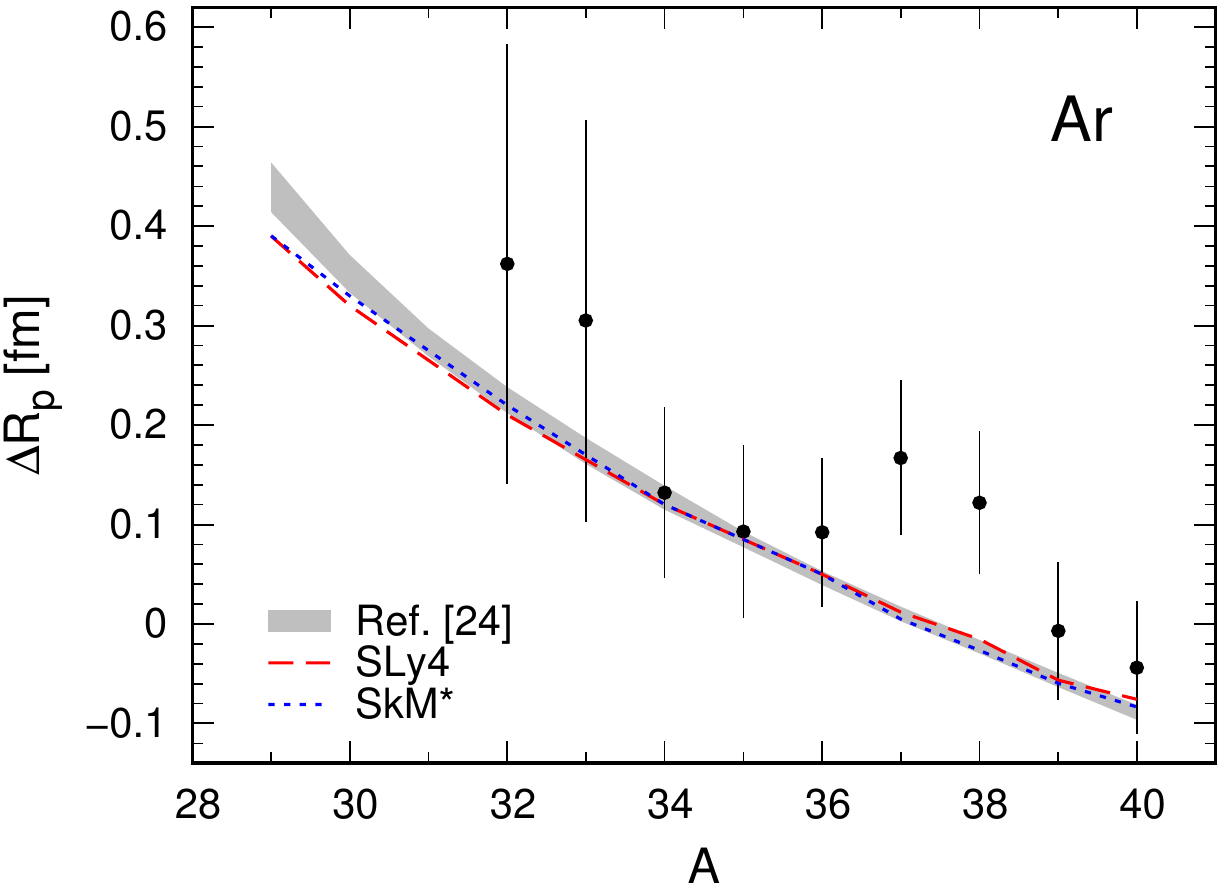}
\caption{(Color online) Predicted proton skins of Ar isotopes as a
function of the mass number $A$. The results calculated with SLy4
and SkM* Skyrme interactions are given by red dashed and blue
dotted lines, respectively. The gray band represents the result of
Sammarruca \cite{Sammarruca2018}. The experimental data points are
from Ozawa {\it et al.} \cite{Ozawa2002}.
\label{fig1}}
\end{figure}

Next, we show in Fig.~\ref{fig1} the calculated HFB results for
the proton skins of Argon isotopes with $A=29-40$. They are
compared with the theoretical predictions from
\cite{Sammarruca2018} and the experimental proton-skin thicknesses
for isotopes $^{32-40}$Ar deduced from the interaction cross
sections of $^{31-40}$Ar on carbon target \cite{Ozawa2002}. In
spite of the large errors of the empirical data, the predictions
of both theoretical methods describe reasonably well their trend,
namely the monotonic decrease of the proton skin with increasing
of the neutron number $N$ in a given isotopic chain. Actually, the
data in Fig.~\ref{fig1} cover a range of $N$ that includes the
magic number $N=20$, but an enhancement of the proton skin is seen
earlier at $N=19$. It was pointed out in Ref.~\cite{ElAdri2019}
that in the case of Argon isotopes $N=20$ was not found to be a
magic number. Most likely this is due to the inversion of the
standard $sd$-shell configuration and the intruder $fp$-shell, as
it has been proved for the neutron-rich $^{32}$Mg nucleus, which
lies in the much explored island of inversion at $N=20$ (see, for
instance, Ref.~\cite{Gaidarov2014}). Obviously, more detailed
consideration of the Ar isotopes around $A=38$ is necessary to
give a clear answer about the role of the shell effects on the
behavior of the proton skin data in this mass region.

\begin{figure}
\centering
\includegraphics[width=1.00\linewidth]{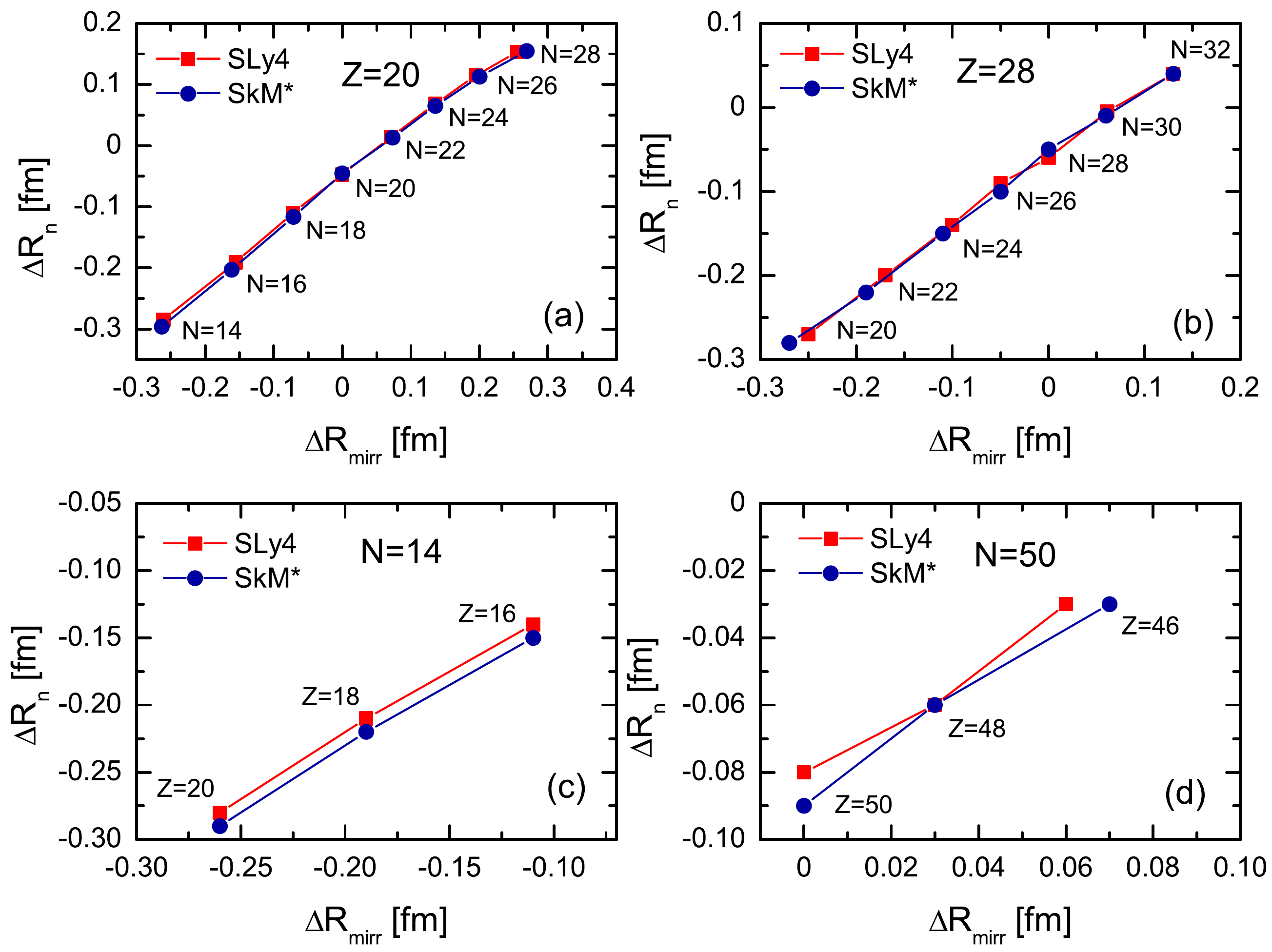}
\caption{(Color online) Relation between the neutron skin $\Delta
R_{n}$ and $\Delta R_{mirr}$ [Eq.~(\ref{eq:6})] for the $Z=20$ (a)
and $Z=28$ (b) isotopic chains and for the N=14 (c) and N=50 (d)
isotonic chains. The results with SLy4 and SkM* forces are given
by red (with squares) and blue (with dots) lines, respectively.
\label{fig2}}
\end{figure}

The relation between the neutron skin $\Delta R_{n}$ and the
proton radii difference of the mirror pair $\Delta R_{mirr}$ as
defined in Eq.~(\ref{eq:6}) is presented in Fig.~\ref{fig2} on the
examples of $Z=20$ and $Z=28$ isotopic chains and two isotonic
chains with $N=14$ and $N=50$. In all four cases a linear relation
between these characteristics is observed. Note that the results
are obtained in the presence of Coulomb effects. It is seen from
Fig.~\ref{fig2} that both Skyrme interactions provide similar
results. To explore to which extent the linear relation between
$\Delta R_{n}$ and $\Delta R_{mirr}$
\begin{equation}
\Delta R_{n}=c(\Delta R_{mirr})+d
\label{eq:21}
\end{equation}
holds based on the four chains, we perform a linear fit of the
curves corresponding to SkM* and SLy4 forces. The parameters of
the generalized linear relation are the following:
\begin{equation}
c=0.866\pm 0.037 ,  \;\;\;\;   d=-0.0633\pm 0.0041
\label{eq:22}
\end{equation}
in the case of SkM* force and
\begin{equation}
c=0.862\pm 0.042 ,   \;\;\;\;  d=-0.0575\pm 0.0041
\label{eq:23}
\end{equation}
in the case of SLy4 force. We would like to note that the
exhibited linear relation does not depend much on the effective
Skyrme interaction used in the calculations. Moreover, based on
microscopic EOS similar linear relations are shown in
Ref.~\cite{Sammarruca2018} for chains with $N=28$, $Z=10$, and
$Z=20$. Confirming a global relation between $\Delta R_{n}$ and
$\Delta R_{mirr}$ regardless $Z$ and $N$, the two different
theoretical methods used in the present work and in
Ref.~\cite{Sammarruca2018} yield similar predictions of the linear
relation (see Eq.~(11) in \cite{Sammarruca2018}). Thus, measuring
the proton radii of the mirror pair one can get an access to the
neutron-skin thickness of the $(Z,N)$ nucleus
[Eqs.~(\ref{eq:21})-(\ref{eq:23})].

Here we would like to stress that the linear fit performed in the
present calculations follows the line in
Ref.~\cite{Sammarruca2018}, where a {\it family of mirror pairs}
is considered to determine the relation between $\Delta R_{n}$ and
$\Delta R_{mirr}$ for a single interaction, while in Brown
\cite{Brown2017} this correlation is determined for a {\it
particular mirror pair} ($^{52}$Ni--$^{52}$Cr) using 48 different
Skyrme functionals. Also, as was pointed out in
Ref.~\cite{Sammarruca2018}, Eqs.~(\ref{eq:21})-(\ref{eq:23}) give
a small negative value of $\Delta R_{n}$ in the limit of $\Delta
R_{mirr}=R_{p}(N,Z)-R_{p}(Z,N)\rightarrow 0$, which makes sense in
the light of Coulomb effects. Hence, Eq.~(\ref{eq:6}) suggests
appropriate modifications to account for Coulomb effects including
the slope $c<1$. As a result, a small deviation from the obtained
linear fit of the relation (\ref{eq:21}) can be observed for the
case of the mass region of heavier $N=50$ isotones [given in
Fig.~\ref{fig2}(d)].

\subsection{Symmetry energy of nuclei from isotopic chains with
$Z=20$ and $Z=28$ and isotonic chains with $N=14$ and $N=50$ and
their respective mirror partner nuclei}

In what follows, we show our results for the symmetry energy $S$
obtained within the CDFM using first the Brueckner EDF for the
symmetry energy in infinite nuclear matter $S^{ANM}$ in
Eq.~(\ref{eq:18}), while the weight function $|{\cal F}(x)|^{2}$
is obtained using Eq.~(\ref{eq:16}) by means of the density
distribution within the Skyrme HFB method (the HFBTHO densities).
Second, we calculate in the CDFM the symmetry energy $S$ using
also the Skyrme EDF. In this case there is a self-consistency
between the way to obtain $|{\cal F}(x)|^{2}$ in the Skyrme HFB
method and the use of the Skyrme EDF to obtain the symmetry
energy. Also, we inspect the correlation of the neutron-skin
thickness $\Delta R_{n}$ of nuclei in a given isotopic chain with
the $S$ [Eq.~(\ref{eq:18})], $p_{0}$ [Eq.~(\ref{eq:19})], and
$\Delta K$ [Eq.~(\ref{eq:20})] parameters extracted from the
density dependence of the symmetry energy around the saturation
density. Complementary to the analyses in
Refs.~\cite{Gaidarov2011,Gaidarov2012}, in the present work we put
emphasis on the possible existence of such correlations for the
chain of mirror partners of the corresponding isotopes, which is
one of the main tasks of the present work.

\begin{figure}
\centering
\includegraphics[width=0.48\linewidth]{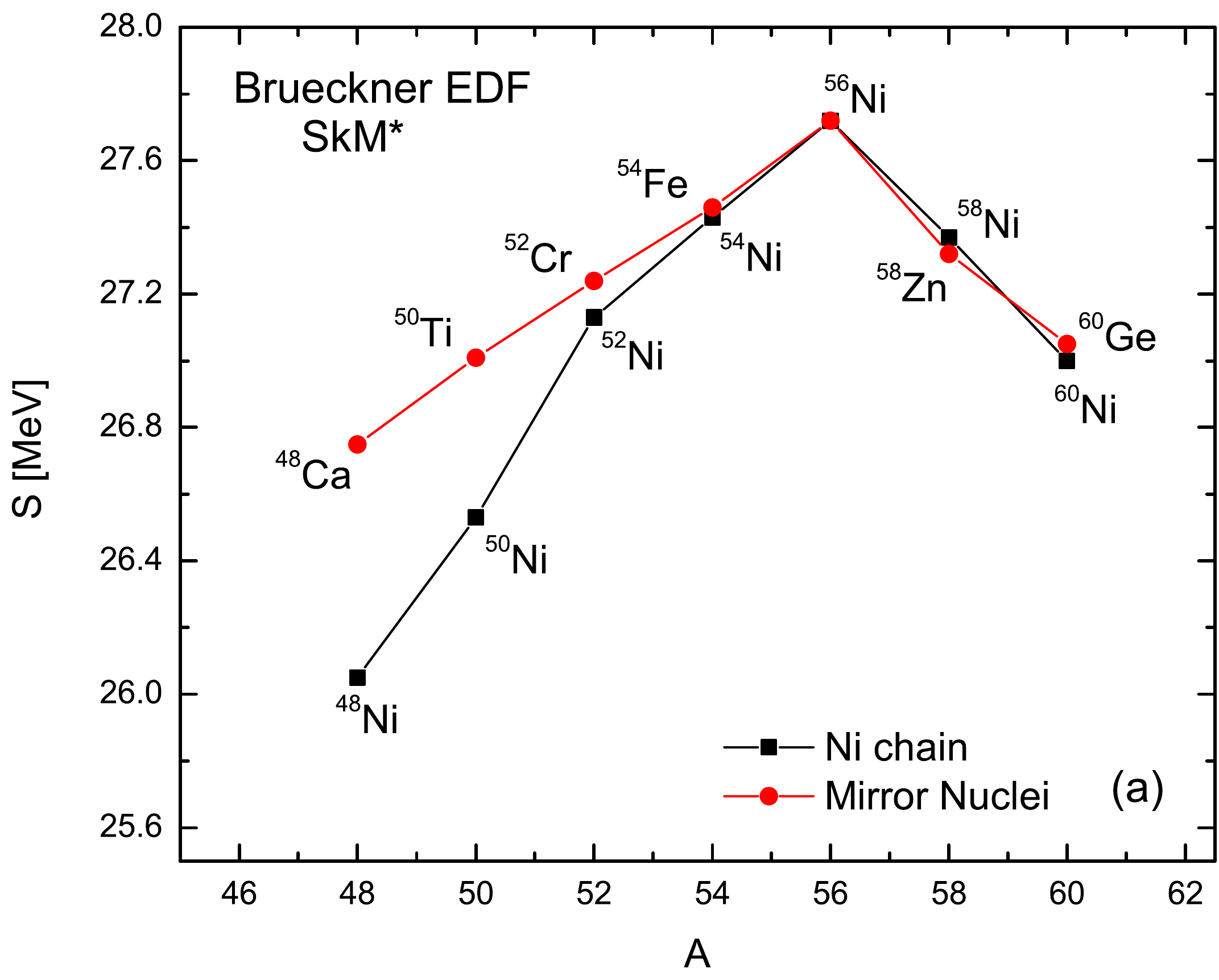}
\includegraphics[width=0.48\linewidth]{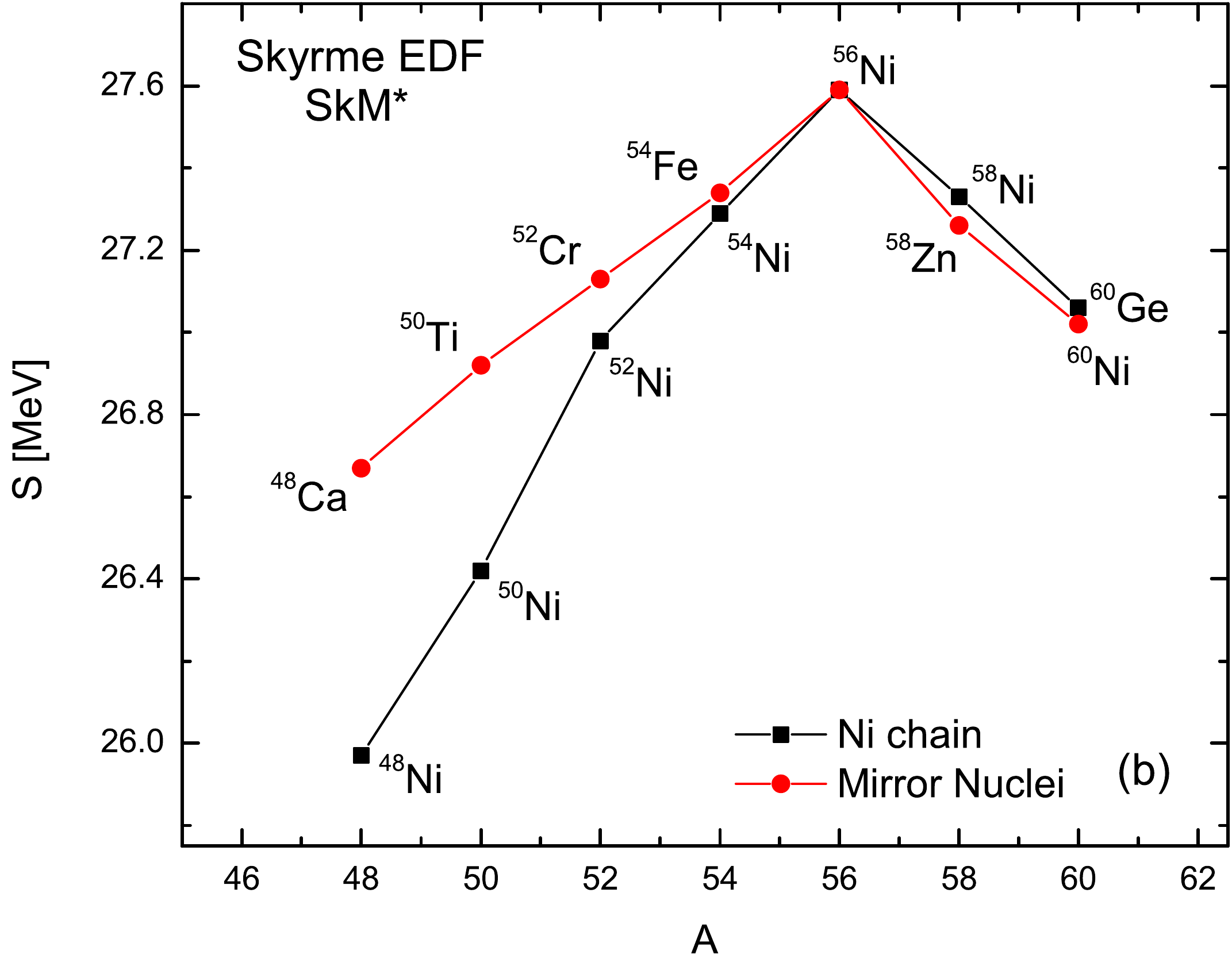}
\vspace{.3cm}\\
\includegraphics[width=0.48\linewidth]{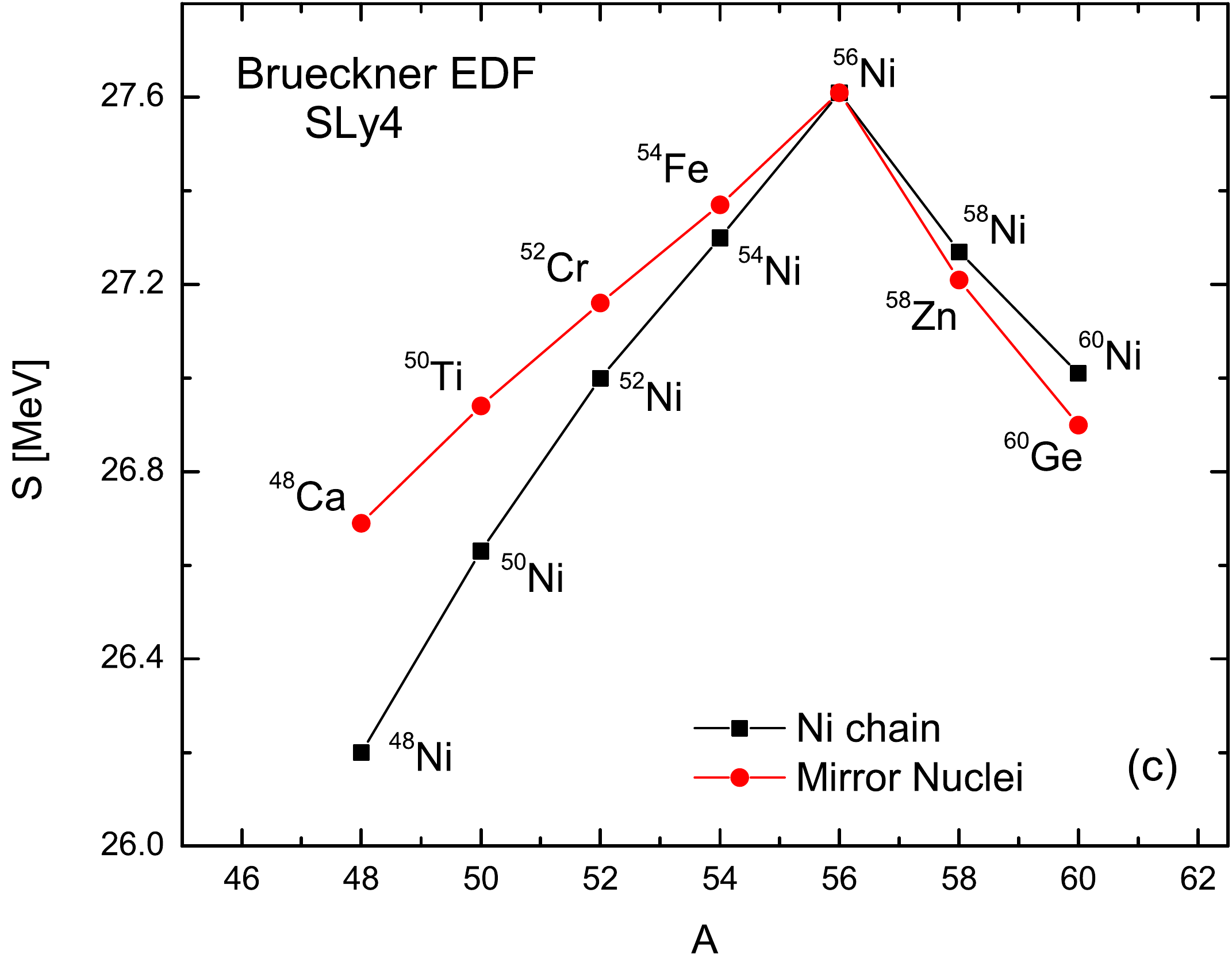}
\includegraphics[width=0.48\linewidth]{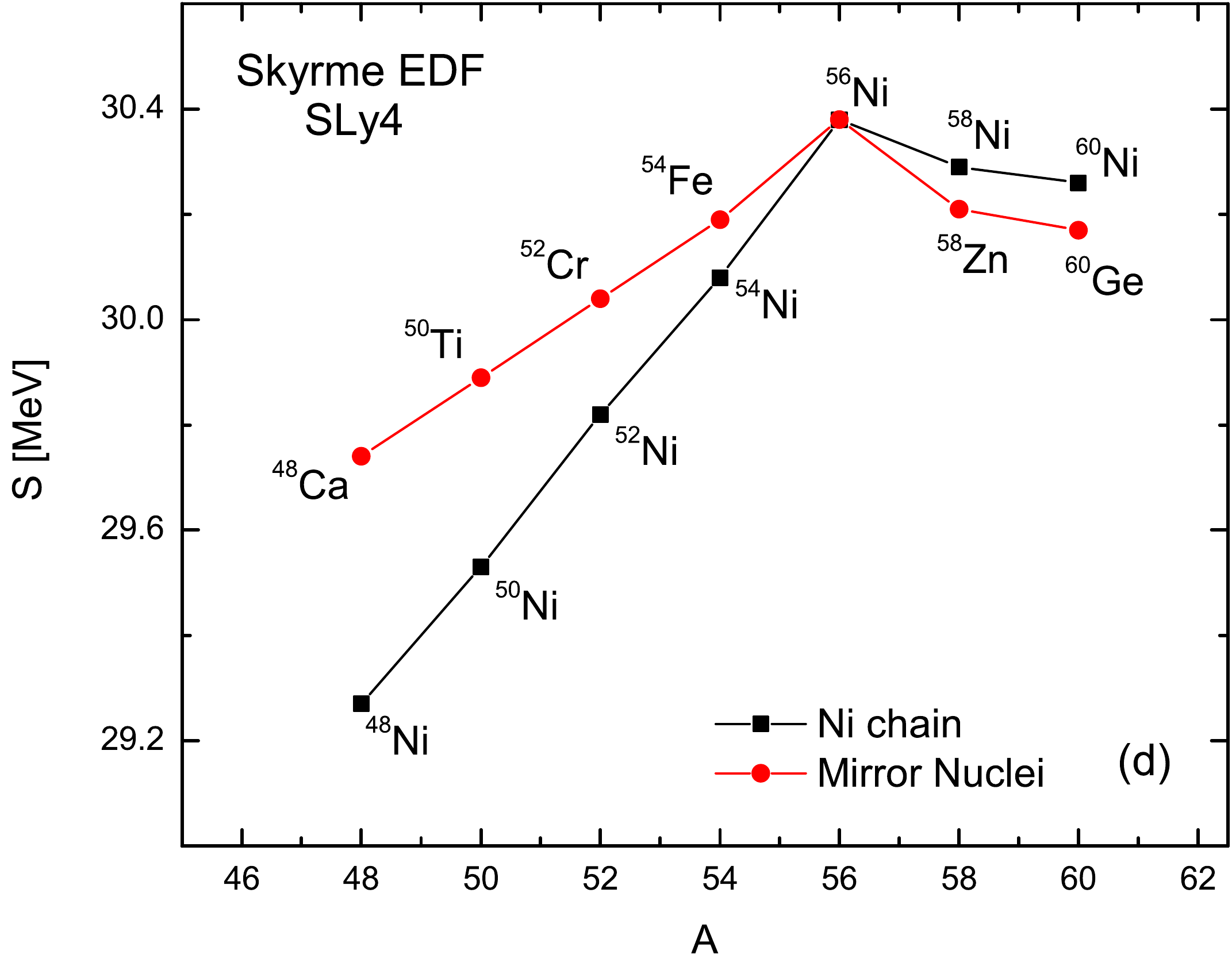}
\caption{(Color online) The symmetry energy $S$ as a function of
the mass number $A$ for Ni isotopes (black line with squares) and
their mirror nuclei (red line with circles) calculated with
Brueckner [panels (a) and (c)] and Skyrme [panels (b) and (d)]
EDFs in the case of SkM* and SLy4 forces, respectively.
\label{fig3}}
\end{figure}

The mass dependence of the symmetry energy $S$ for the Ni isotopes
($Z=28$) with $A=48-60$ and their mirror nuclei by using the
Brueckner and Skyrme EDFs with SLy4 and SkM* forces is presented
in Fig.~\ref{fig3}. The behavior of the curves for the two
functionals is similar and the values of $S$ obtained with Skyrme
EDF with SLy4 force are larger in comparison with the
corresponding values deduced in other cases.
An important result, which can be seen from Fig.~\ref{fig3}, is
that the mirror partner nuclei show the same linear behavior
observed in the evolution of the symmetry energy in Ni chain (see
also Ref.~\cite{Gaidarov2011}) containing an inflection point
("kink") at the double-magic $^{56}$Ni nucleus. We note that there
is a small shift of the curve for mirror partners with respect to
the curve for the Ni isotopes with a smooth growing deviation
between them with increasing $|N-Z|$.

\begin{figure}
\centering
\includegraphics[width=0.5\linewidth]{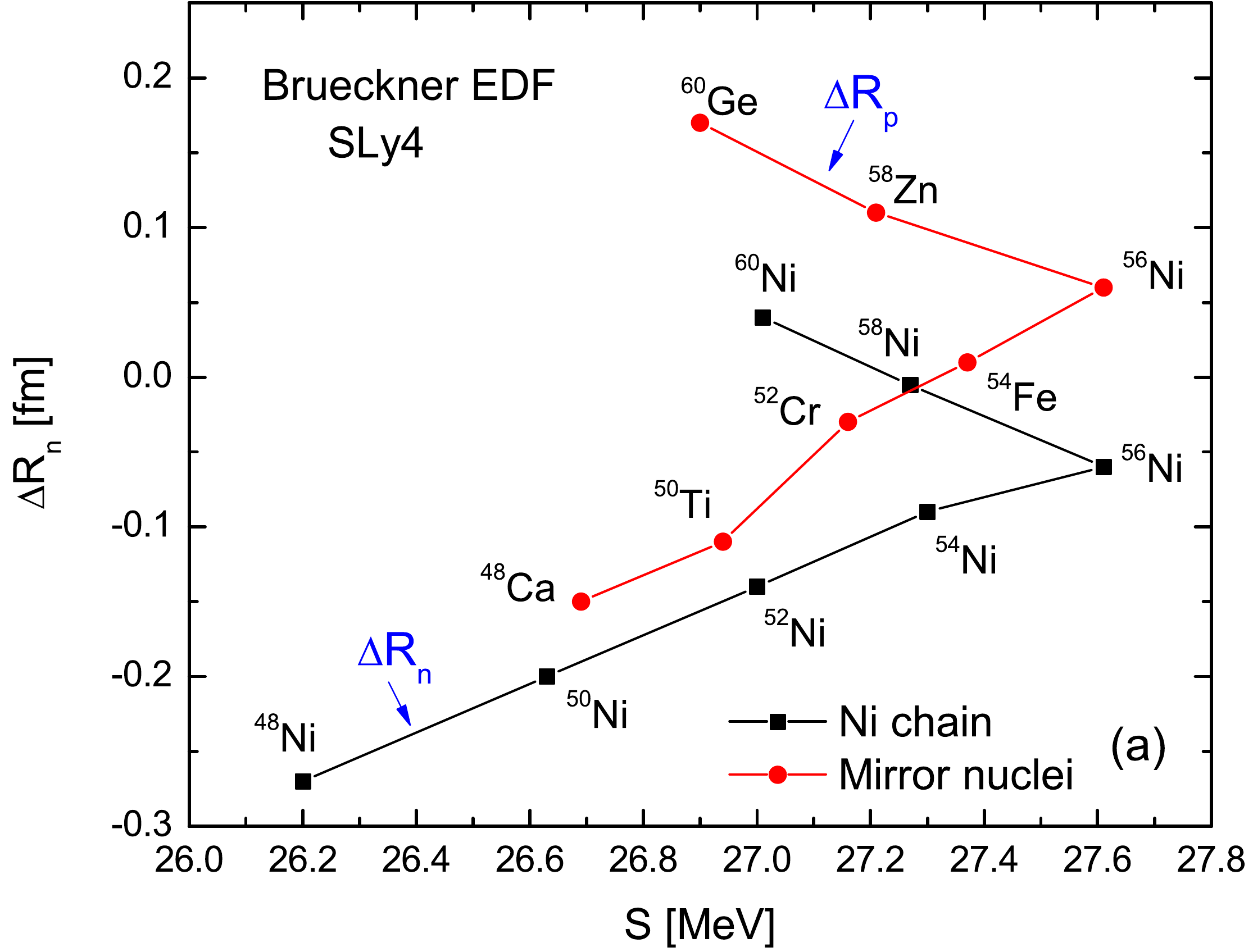}
\includegraphics[width=0.48\linewidth]{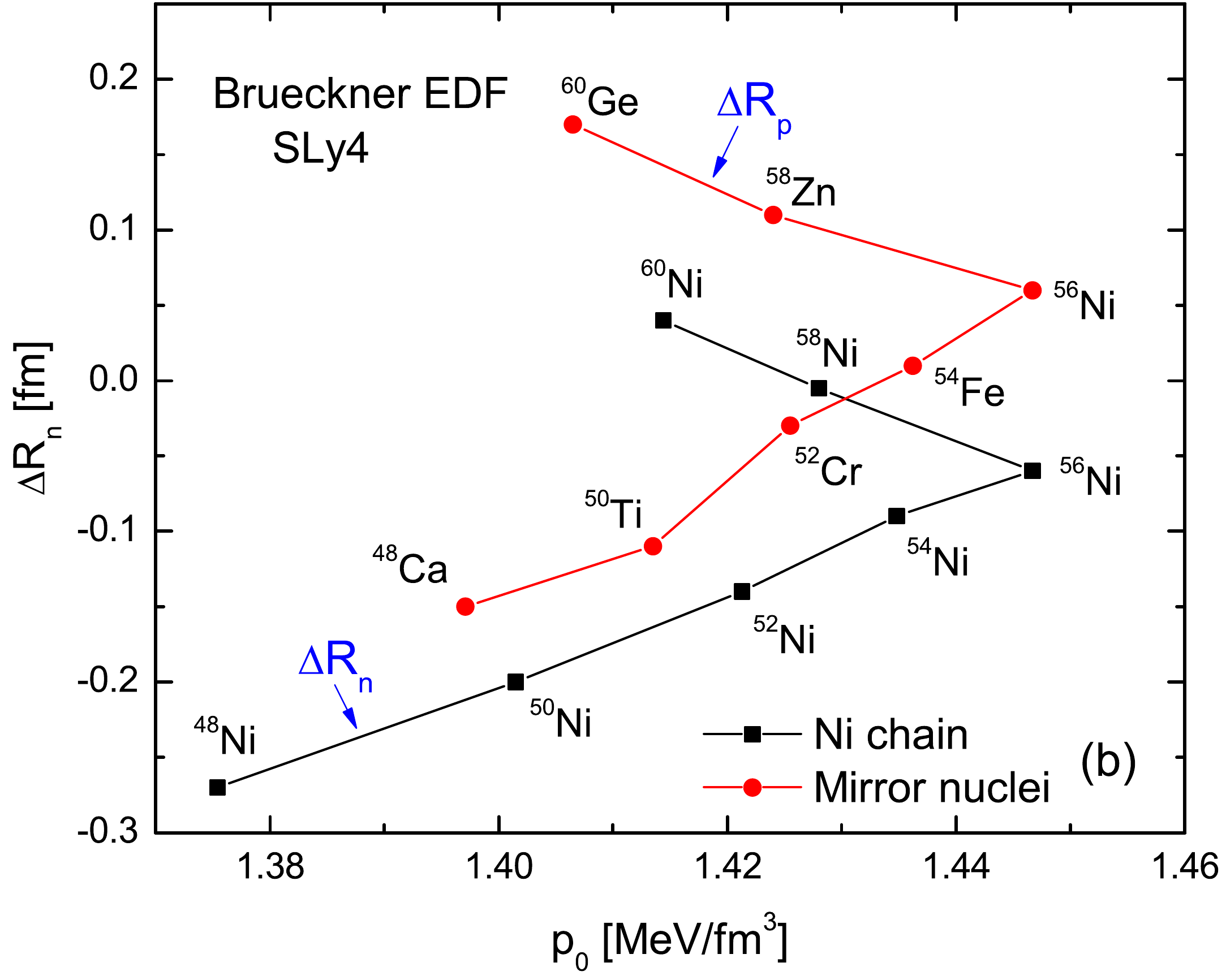}
\vspace{.3cm}\\
\includegraphics[width=0.49\linewidth]{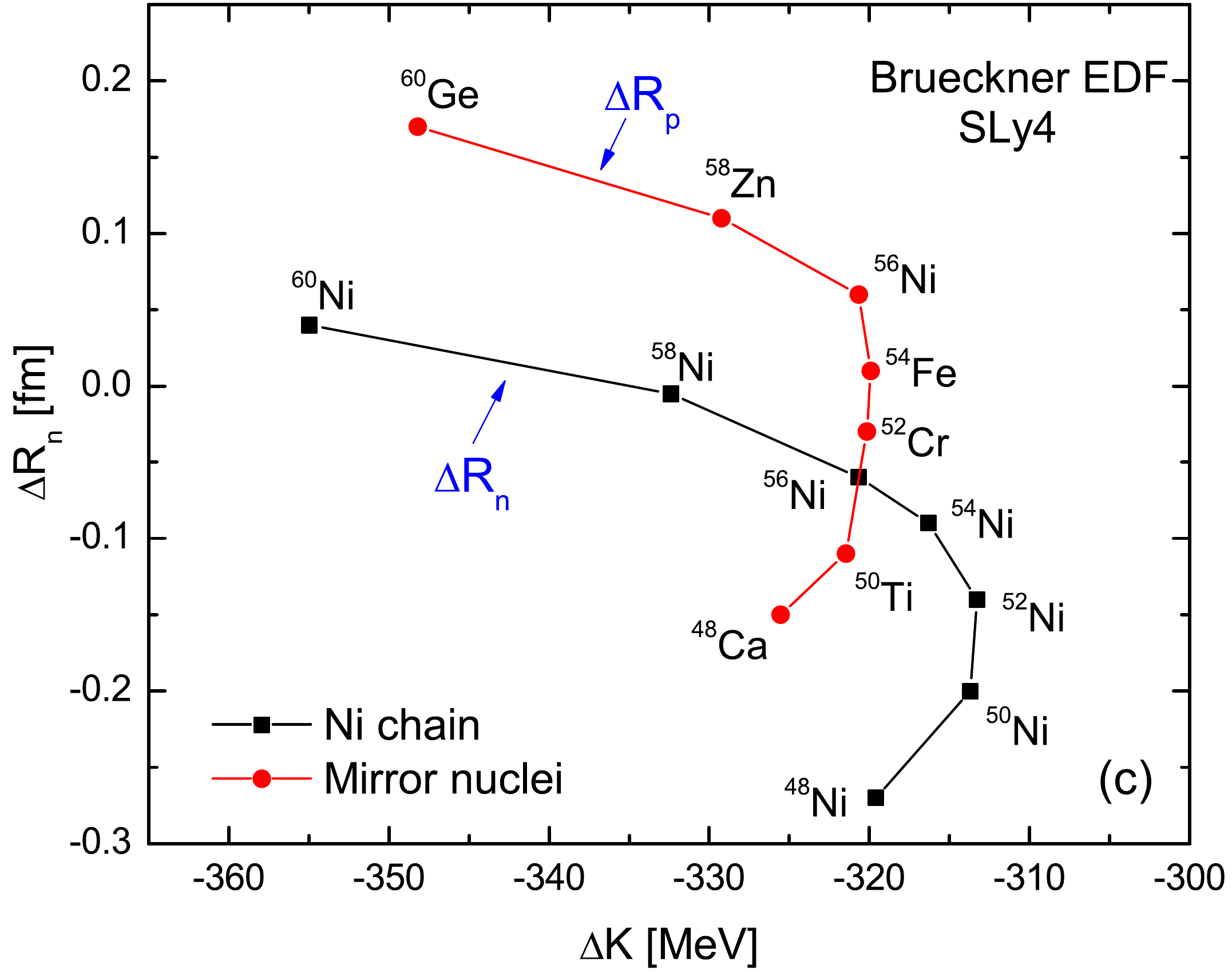}
\caption{(Color online) HFB neutron-skin thickness $\Delta R_{n}$
for Ni isotopes (black line with squares) and corresponding proton
skins $\Delta R_{p}=-\Delta R_{n}$ for their mirror nuclei (red
line with circles) as a function of the symmetry energy $S$ (a),
pressure $p_{0}$ (b), and asymmetric compressibility $\Delta K$
(c) calculated with Brueckner EDF and SLy4 force. \label{fig4}}
\end{figure}

The results for the correlation between the neutron(proton)-skin
thickness and the symmetry energy, the pressure, and the
asymmetric compressibility for the Ni isotopes and their mirror
partners calculated on the basis of the Brueckner EDF for ANM and
using SLy4 force are shown in Fig.~\ref{fig4}. It is seen from the
figure that there exists an approximate linear correlation between
$\Delta R_{n}$ and $S$ not only for the even-even Ni isotopes with
$A=48-60$, but also for the mirror nuclei between $\Delta
R_{p}=-\Delta R_{n}$ and $S$. We observe a smooth growth of the
symmetry energy till the double-magic nucleus $^{56}$Ni and then a
linear decrease of $S$ while the neutron-skin thickness of the
isotopes increases. This tendency happens also for the chain of
the mirror nuclei from $^{48}$Ca to $^{60}$Ge with an inflection
point transition at the double-magic $^{56}$Ni nucleus. The
correlation between $\Delta R_{n}$ and $p_{0}$ is similar, while a
less strong correlation between $\Delta R_{n}$ and $\Delta K$ is
found. Here we note that the determined values of the neutron-skin
thickness $\Delta R_{n}$ of $^{48}$Ca nucleus are 0.155 fm in the
case of SkM* and 0.154 fm in the case of SLy4 forces,
respectively, in consistency with the predicted value of 0.159 fm
using the Gogny-D1S HFB method \cite{Tagami2020}. These values are
also in good agreement with {\it ab initio} calculations based on
the chiral effective field theory that yield values between 0.14
fm and 0.20 fm for the neutron skin of $^{48}$Ca (see, for
instance, the value of $0.181\pm 0.010$ in
Ref.~\cite{Sammarruca2018}).

\begin{figure}[h]
\centering
\includegraphics[width=0.48\linewidth]{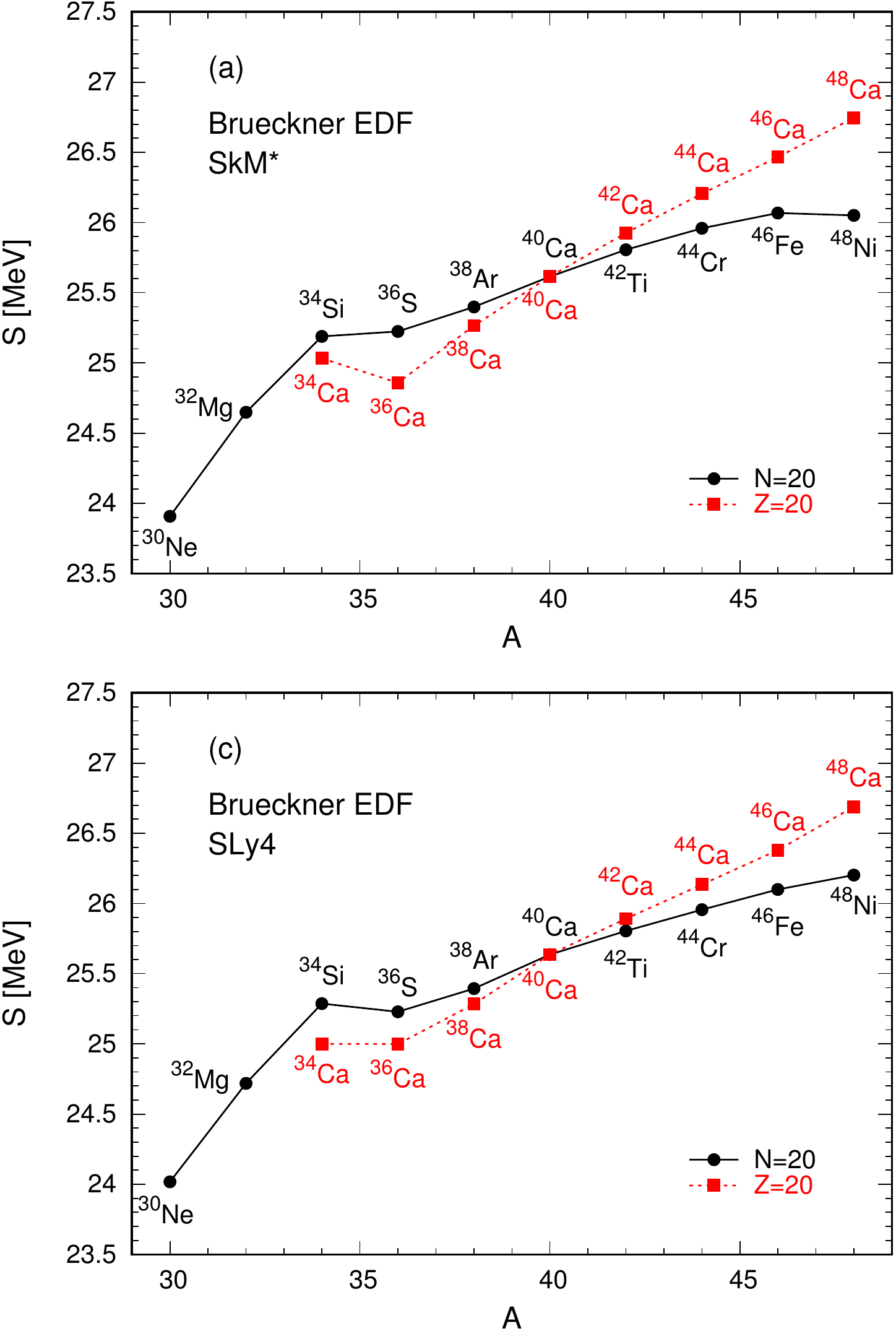}
\includegraphics[width=0.48\linewidth]{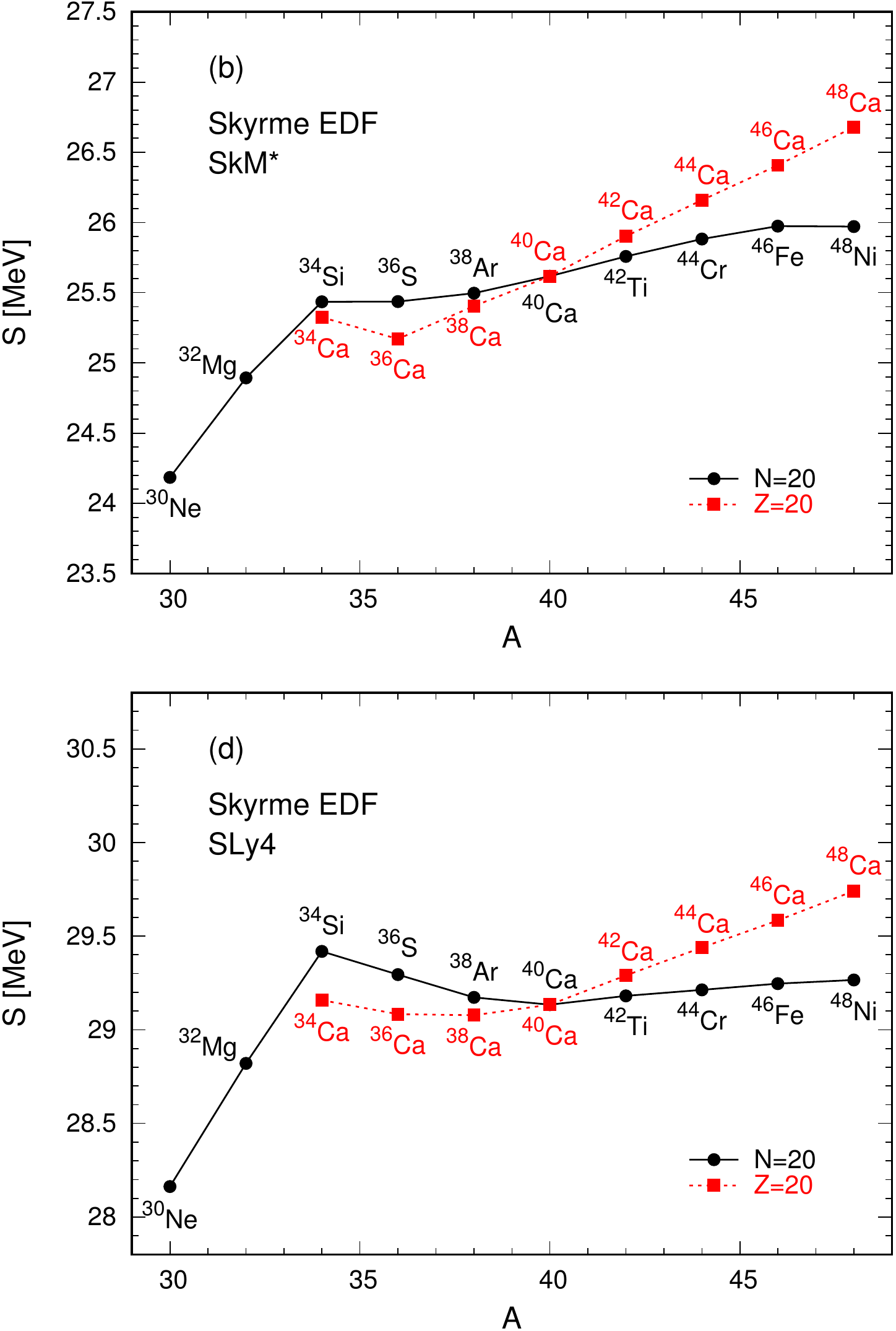}
\caption{(Color online) The symmetry energy $S$ as a function of
the mass number $A$ for Ca isotopes (red dotted line with squares)
and their mirror nuclei (black solid line with circles) calculated
with Brueckner [panels (a) and (c)] and Skyrme [panels (b) and
(d)] EDFs in the case of SkM* and SLy4 forces, respectively.
\label{fig5}}
\end{figure}

In Fig.~\ref{fig5} we display results on the isotopic evolution of
the symmetry energy for the $Z=20$ isotopes $(A=34-48)$ and their
corresponding mirror partners, extending the latter to include two
more isotones with $Z=10$ and 12. They are shown for the cases of
Brueckner and Skyrme EDFs with both SkM* and SLy4 forces. This
analysis is motivated by the active study of Ca isotopes in the
proximity of $^{48}$Ca, with emphasis on the evolution of the
charge radii at both proton-rich \cite{Miller2019} and
neutron-rich \cite{Ruiz2016} sides. As seen in Fig.~\ref{fig5},
our results for the symmetry energy $S$ exhibit a characteristic
monotonic dependence on $A$ for $A>36$, in accordance with the
predictions for the charge radii of the Ca isotopes calculated
within the HFB method by using of the Skyrme parametrization
SV-min \cite{Miller2019}. In contrast to the Ni chain presented in
Fig.~\ref{fig3} where a "kink" exists at the double-magic
$^{56}$Ni, no "kink" is observed, particularly at the double-magic
$^{40}$Ca nucleus. However, there is a "kink" at $^{36}$Ca that is
more pronounced when SkM* force is used in the calculations with
both Brueckner and Skyrme EDFs. The corresponding mirror partners
($N=20$ isotones) also reveal a linear behavior but with an
inflection point transition at $^{34}$Si nucleus that is more
pronounced with SLy4 force.

The isotopic(isotonic) evolution of the symmetry energy of nuclei
with $Z(N)=20$ shown in Fig.~\ref{fig5} requires a deeper
analysis. The existence of kinks in the symmetry energy behavior
in these chains at $^{36}$Ca and $^{34}$Si (mostly pronounced with
the use of SLy4 interaction) can be attributed to the specific
nuclear shell structure in this mass region. Based on the
evolution of neutron separation energies and cross sections in
light nuclei, it was originally proposed $N=16$ to be a new magic
number lying between the usual $N=8$ and 20 for N, O and F nuclei
\cite{Ozawa2000}. It has been observed experimentally
\cite{Warburton90} that the $N=20$ isotonic chain shows an
evidence of two new sub-shell closures for $N=14$ and $N=16$. In
fact, $^{34}$Si and $^{36}$S reveal the typical features of
double-magic nuclei. The predicted double-magicity of $^{34}$Si is
discussed in \cite{Angeli2015} but, to our knowledge, no
unambiguous evidence has been found yet. Nevertheless, the
reported first experimental proof that points to a depletion of
the central density of protons in the short-lived nucleus
$^{34}$Si shown in Ref.~\cite{Mutschler2017} is in favour of the
double-magicity structure of this nucleus, in which the mixing
between normally occupied and valence orbits is very limited. This
newly observed $N=14$, $N=16$ or $Z=14$, $Z=16$ shell
stabilization in $Z(N)=20$ chains, correspondingly, is expected to
be symmetric with respect to the isospin projection. Therefore,
the ideal region to check such prediction is along the $Z=20$ Ca
isotopes and their mirror $N=20$ isotones. The detailed
spectroscopy of $^{36}$Ca and its mirror nucleus $^{36}$S made
with RISING at GSI led to the hypothesis of a new "island of
inversion" developed by neutron-deficient Ca isotopes where the
onset of inversion may start already at $N=14$ in $^{34}$Ca
\cite{Doornenbal2007}.

\begin{figure}
\centering
\includegraphics[width=0.48\linewidth]{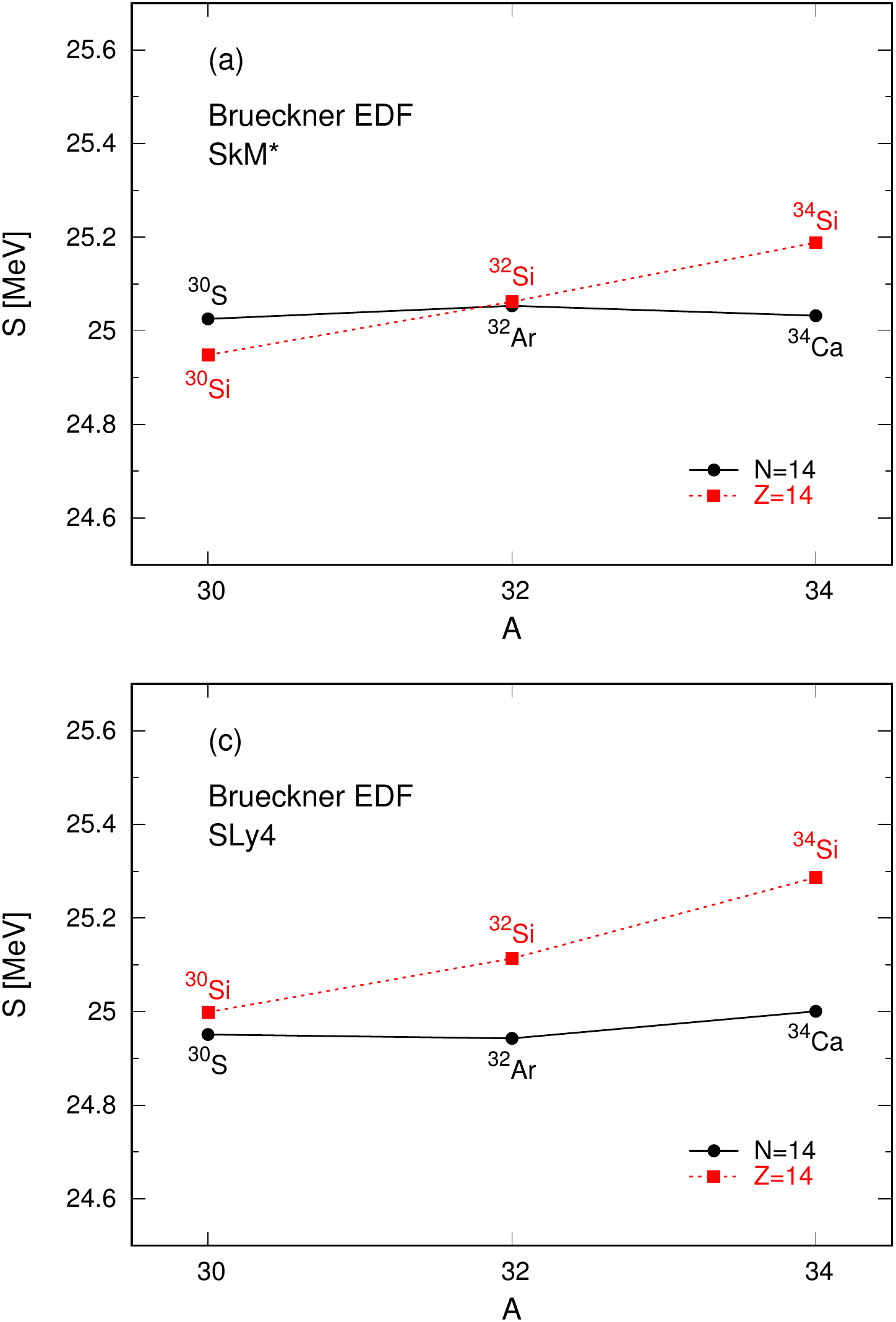}
\includegraphics[width=0.48\linewidth]{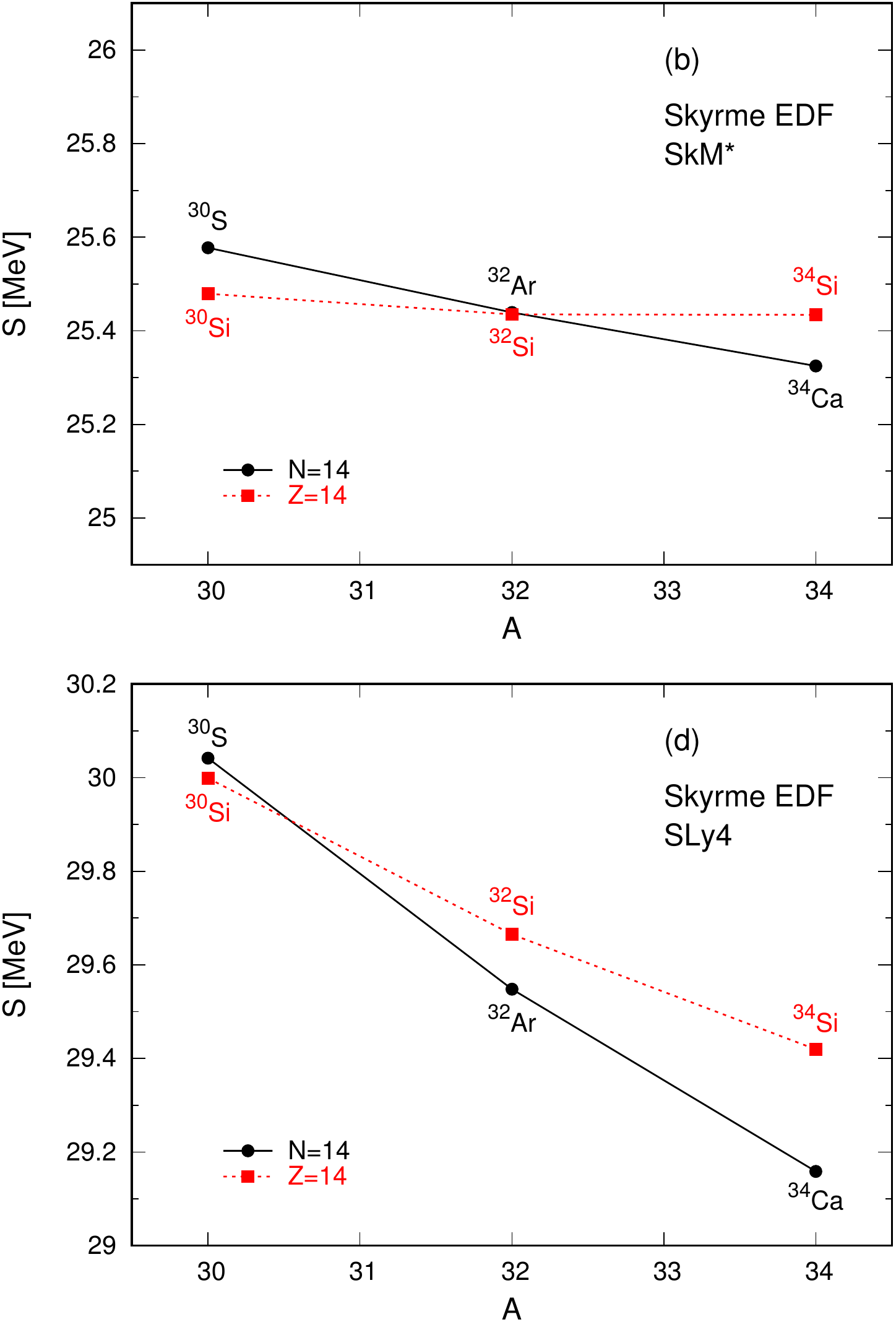}
\caption{(Color online) The symmetry energy $S$ as a function of
the mass number $A$ for $N=14$ isotones (black solid line with
circles) and their mirror Si isotopes (red dotted line with
squares) calculated with Brueckner [panels (a) and (c)] and Skyrme
[panels (b) and (d)] EDFs in the case of SkM* and SLy4 forces,
respectively. \label{fig6}}
\end{figure}

Figure \ref{fig6} illustrates the evolution of the symmetry energy
for $N=14$ isotones and their mirror Si isotopes ($Z=14$). Our
choice of this isotonic chain is motivated by the fact that data
for light nuclei, such as those with $N=14$, are likely to be
obtained in future electron scattering facilities such as SCRIT
and ELISe at FAIR in GSI. It is seen from Fig.~\ref{fig6} that the
symmetry energies deduced from SkM* and SLy4 parametrizations vary
in the interval 25--30 MeV.

\begin{figure}
\centering
\includegraphics[width=0.48\linewidth]{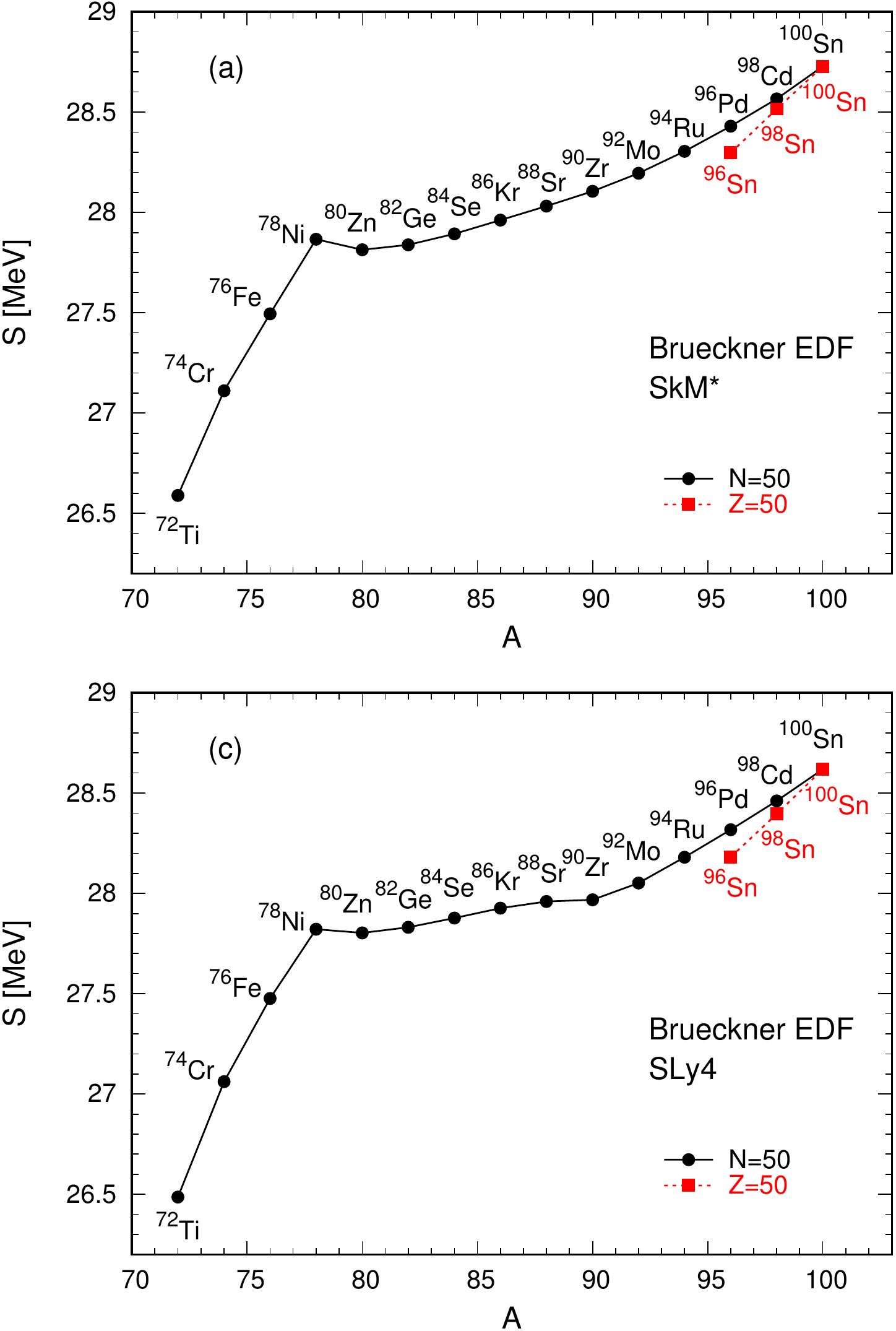}
\includegraphics[width=0.48\linewidth]{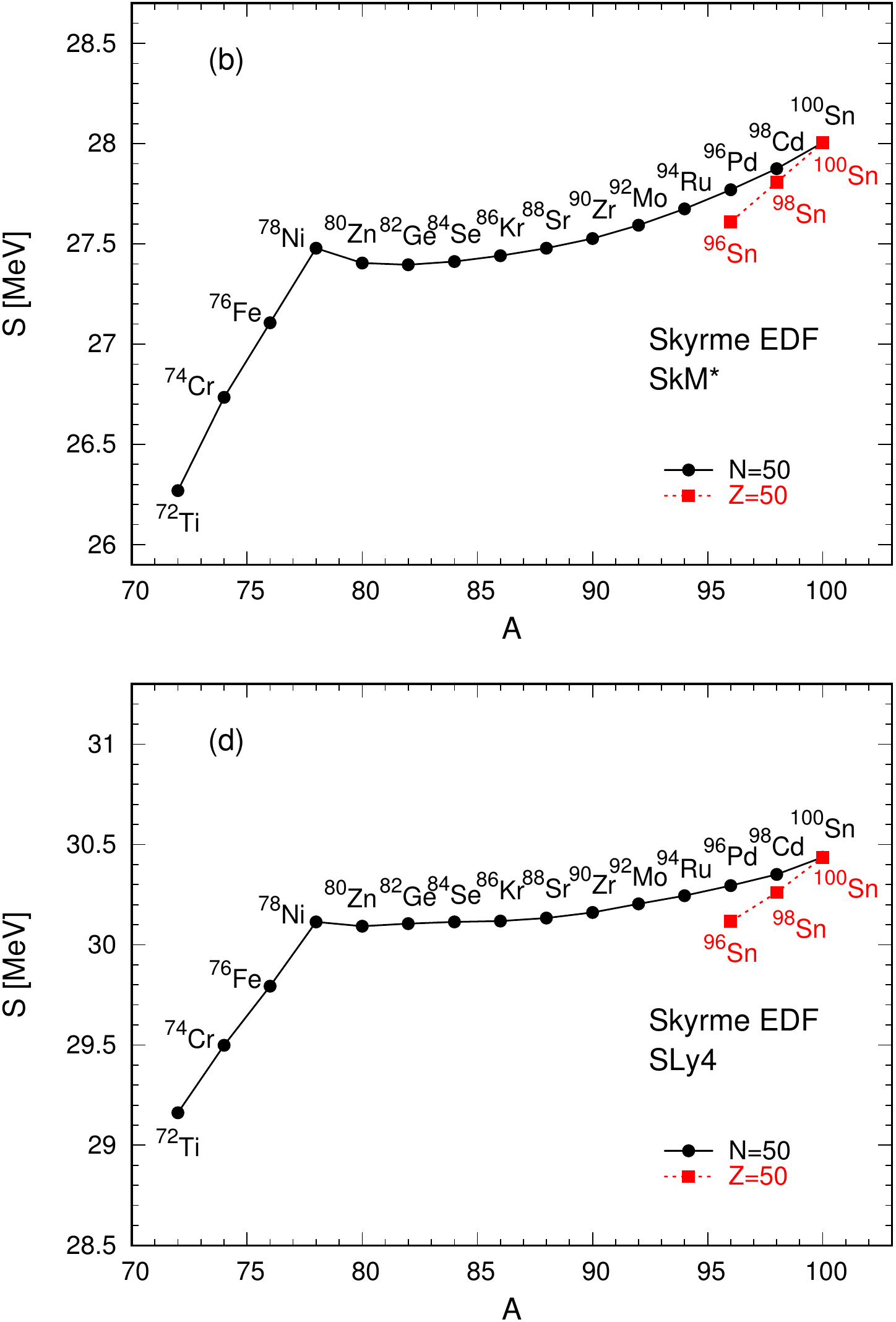}
\caption{(Color online) The symmetry energy $S$ as a function of
the mass number $A$ for $N=50$ isotones (black solid line with
circles) and their heaviest three mirror Sn isotopes (red dotted
line with squares) calculated with Brueckner [panels (a) and (c)]
and Skyrme [panels (b) and (d)] EDFs in the case of SkM* and SLy4
forces, respectively. \label{fig7}}
\end{figure}

Finally, we show in Fig.~\ref{fig7}, in the same manner as in
Figs.~\ref{fig3}, \ref{fig5} and \ref{fig6}, the mass dependence
of the nuclear symmetry energy $S$ of even-even nuclei from the
$N=50$ chain $(A=72-100)$ and their heaviest three mirror Sn
isotopes. In this case the need to consider mirror pairs limits
the spectrum of realistic possibilities. First, we can observe a
strong "kink" at the double-magic $^{78}$Ni nucleus along the
chain from $^{72}$Ti to $^{100}$Sn, which is a common feature for
both EDFs and Skyrme forces used in the calculations. Second, the
two curves approach each other going to the double-magic
$^{100}$Sn nucleus, where they are crossing, similarly to the
cases of $Z=28$ and $Z=20$ isotopic chains and their mirror nuclei
presented in Figs.~\ref{fig3} and \ref{fig5}, respectively, where
the intersection takes place at the double-magic $^{56}$Ni and
$^{40}$Ca nuclei.

As a common feature of the results for the nuclear symmetry energy
in different chains and their mirror nuclei presented in
Figs.~\ref{fig3}-\ref{fig7}, the calculations with Skyrme EDF
using SLy4 force yield larger values of $S$ in comparison with
other cases. Similar trend has been found in
Ref.~\cite{Antonov2017}, where the values of the symmetry energy
coefficient $e_{sym}$ for Ni, Sn, and Pb isotopic chains at zero
temperature calculated with Skyrme EDF and SLy4 overestimate those
obtained with the SkM* force. The difference in the magnitude of
the symmetry energy when using both Skyrme parametrizations is
associated with their different saturation properties, namely, for
the parameter set SkM* and SLy4, the corresponding symmetry energy
of nuclear matter $a_{4}$ at saturation density $\rho_{0}$ is
30.00 MeV and 31.99 MeV, respectively (see, for instance,
Ref.~\cite{Wang2015}).

\section{Summary and conclusions \label{s:conclusions}}

In this work, the HFB method by using the cylindrical transformed
deformed harmonic-oscillator basis has been applied to
calculations of radii and skins for several mirror pairs in the
middle mass range. The existence of an important possible
correlation between the neutron(proton) skins and the parameters
of the EOS, such as the symmetry energy $S$, the pressure $p_{0}$,
and the curvature $\Delta K$, has been investigated. The mentioned
EOS parameters have been calculated for Ni isotopic chain with
mass number $A=48-60$, as well as for nuclei with $Z=20$, $N=14$,
and $N=50$ and their respective mirror partner nuclei, using
Brueckner and Skyrme EDFs for isospin ANM with two Skyrme-type
forces, SkM* and SLy4. The results are obtained within the CDFM
that links the properties of nuclear matter with the microscopic
description of finite nuclei.

The main results of the present study can be summarized, as
follows:

i) Due to Coulomb effects, the predicted proton skins are found
larger than the neutron skins of the corresponding mirror partner
nuclei. They compare reasonably well with the available empirical
data, for instance for Ar isotopes, that are also well described
by chiral effective field theory-based EOS;

ii) The studied relation between the neutron skin $\Delta R_{n}$
and the difference between the proton radii $\Delta R_{mirr}$ for
a family of mirror pairs in the presence of Coulomb effects shows
clearly a linear dependence. Thus, this appears to be an
alternative way to explore neutron skins that may challenge
experimentalists to perform high-precision measurements of charge
radii of unstable neutron-rich isotopes;

iii) We found in the case of the Ni isotopic chain and the
respective mirror partner nuclei a strong correlation between the
neutron (proton)-skin thickness and the symmetry energy $S$ and
pressure $p_{0}$ with a "kink" at double-magic $^{56}$Ni, while
the correlation between $\Delta R_{n}$ ($\Delta R_{p}$) and the
asymmetric compressibility $\Delta K$ is less pronounced. In our
opinion, more general conclusions for such correlations (including
also other nuclear chains of mirror pairs) can be drawn after
detailed treatment of nuclear deformation, shell structure and
surface effects;

iv) The evolution of the symmetry energy $S$ with the mass number
$A$ of nuclei from $Z=20$, $Z=28$, and $N=50$ chains and their
mirror nuclei exhibits similar behavior. The curves cross in each
chain at the corresponding $N=Z$ nucleus ($^{40}$Ca, $^{56}$Ni,
$^{100}$Sn) and start to deviate from each other with the increase
of the level of asymmetry $|N-Z|$.

In principle, studies of elastic and quasi-elastic electron
scattering on isotopic and isotonic chains can provide useful
information about the evolution of the charge form factors and
related charge density distributions, and thus, on the occupation
and filling of the single-particle levels of nucleons along the
chains (see, e.g.,
Refs.~\cite{Antonov2005,RocaMaza2013,Meucci2014}). These
investigations being combined with analyses of isospin-dependent
properties through the nuclear symmetry energy, its density
dependence and related quantities along chains of mirror pairs,
may lead to observations of new phenomena related to the
proton-to-neutron asymmetry.

\section{Acknowledgments}
M.K.G., A.N.A., and D.N.K. are grateful for the support of the
Bulgarian National Science Fund under Contract No.~KP-06-N38/1.
P.S. acknowledges support from Ministerio de Ciencia, Innovaci\'on
y Universidades MCIU/AEI/FEDER, UE (Spain) under Contract No.
PGC2018-093636-B-I00.


\section*{References}
\begin{thebibliography}{99}

\bibitem{Reinhard2010} P.-G. Reinhard, W. Nazarewicz, Phys. Rev. C 81 (2010) 051303.

\bibitem{Simon2007} H. Simon, Nucl. Phys. A 787 (2007) 102.

\bibitem{Antonov2011} A.N. Antonov {\it et al.}, Nucl. Instr. and Meth. in Phys. Res. A 637 (2011) 60.

\bibitem{Suda2005} T. Suda, M. Wakasugi, Prog. Part. Nucl. Phys. 55 (2005) 417.

\bibitem{Suda2012} T. Suda {\it et al.}, Prog. Theor. Exp. Phys. 2012 (2012) 03C008.

\bibitem{Wakasugi2013} M. Wakasugi {\it et al.}, Nucl. Instr. and Meth. in Phys. Res. B 317 (2013) 668.

\bibitem{Tsukada2017} K. Tsukada {\it et al.}, Phys. Rev. Lett. 118 (2017) 262501.

\bibitem{Moreno2009} O. Moreno, P. Sarriguren, E. Moya de Guerra, J.M. Udias,
T.W. Donnelly, I. Sick, Nucl. Phys. A 828 (2009) 306.

\bibitem{Donnelly89} T.W. Donnelly, J. Dubach, I. Sick, Nucl. Phys. A 503 (1989) 589.

\bibitem{prex} http://hallaweb.jlab.org/parity/prex.

\bibitem{Abrahamyan2012} S. Abrahamyan {\it et al.}, Phys. Rev. Lett. 108 (2012) 112502.

\bibitem{Gaidarov2012} M.K. Gaidarov, A.N. Antonov, P. Sarriguren, E. Moya de Guerra, Phys. Rev. C 85 (2012) 064319.

\bibitem{prexII} P.A. Souder {\it et al.}, PREX-II: Precision parity-violating measurement
of the neutron skin of lead; http://hallaweb.jlab.org/
parity/prex/ (2011).

\bibitem{CREX} J. Mammei {\it et al.}, CREX: Parity-violating measurement of the
weak charge distribution of $^{48}$Ca to 0.02 fm accuracy; http://
hallaweb.jlab.org/parity/prex/ (2013).

\bibitem{Horowitz2014} C.J. Horowitz, K.S. Kumar, R. Michaels, Eur. Phys. J. A 50 (2014) 48.

\bibitem{Furnstahl2002} R. Furnstahl, Nucl. Phys. A 706 (2002) 85.

\bibitem{Mahzoon2017} M.H. Mahzoon, M.C. Atkinson, R.J. Charity, W.H. Dickhoff,
Phys. Rev. Lett. 119 (2017) 222503.

\bibitem{Mahzoon2014} M.H. Mahzoon, R.J. Charity, W.H. Dickhoff, H. Dussan, S.J. Waldecker,
Phys. Rev. Lett. 112 (2014) 162503.

\bibitem{Atkinson2020} M.C. Atkinson, M.H. Mahzoon, M.A. Keim, B.A. Bordelon,
C.D. Pruitt, R.J. Charity, W.H. Dickhoff, Phys. Rev. C 101 (2020)
044303.

\bibitem{Pruitt2020} C.D. Pruitt, R.J. Charity, L.G. Sobotka, M.C. Atkinson, W.H.
Dickhoff, Phys. Rev. Lett. 125 (2020) 102501.

\bibitem{Li2019} H. Li, H. Xu, Y. Zhou, X. Wang, J. Zhao, L.-W. Chen, F. Wang, arXiv:nucl-th/1910.06170.

\bibitem{Brown2017} B.A. Brown, Phys. Rev. Lett. 119 (2017) 122502.

\bibitem{Yang2018} J. Yang, J. Piekarewicz, Phys. Rev. C 97 (2018) 014314.

\bibitem{Sammarruca2018} F. Sammarruca, Front. Phys. 6:90 (2018).

\bibitem{NSE2014} Topical issue on Nuclear Symmetry Energy. Guest editors: Bao-An
Li, Angels Ramos, Giuseppe Verde, Isaac Vida\~{n}a. Eur. Phys. J.
A 50 (2014) 2.

\bibitem{Lattimer2007} J.M. Lattimer, M. Prakash, Phys. Rep. 442 (2007) 109.

\bibitem{Li2008} B.A. Li {\it et al.}, Phys. Rep. 464 (2008) 113.

\bibitem{Typel2001} S. Typel, B.A. Brown, Phys. Rev. C 64 (2001) 027302.

\bibitem{Steiner2005} A. Steiner, M. Prakash, J. Lattimer, P. Ellis, Phys. Rep. 411 (2005) 325.

\bibitem{RocaMaza2011} X. Roca-Maza, M. Centelles, X. Vi\~{n}as, M. Warda, Phys. Rev. Lett. 106 (2011) 252501.

\bibitem{Gaidarov2011} M.K. Gaidarov, A.N. Antonov, P. Sarriguren, E. Moya de Guerra, Phys. Rev. C 84 (2011) 034316.

\bibitem{Ant80} A.N.~Antonov, V.A.~Nikolaev, I.Zh.~Petkov,
Bulg. J. Phys. 6 (1979) 151; Z. Phys. A 297 (1980) 257; {\it ibid}
304 (1982) 239; Nuovo Cimento A 86 (1985) 23; A.N.~Antonov {\it et
al.}, {\it ibid} 102 (1989) 1701; A.N. Antonov, D.N. Kadrev, P.E.
Hodgson, Phys. Rev. C 50 (1994) 164.

\bibitem{AHP} A~N.~Antonov, P.E.~Hodgson, I.Zh.~Petkov, Nucleon Momentum and Density Distributions in Nuclei, Clarendon
Press, Oxford, 1988; Nucleon Correlations in Nuclei,
Springer-Verlag, Berlin-Heidelberg-New York, 1993.

\bibitem{Gaidarov2014} M.K. Gaidarov, P. Sarriguren, A.N. Antonov, E. Moya de Guerra, Phys. Rev. C 89 (2014) 064301.

\bibitem{Ozawa2002} A. Ozawa {\it et al.}, Nucl. Phys. A 709 (2002) 60.

\bibitem{Stoitsov2013} M.V. Stoitsov {\it et al.}, Comput. Phys. Commun. 184 (2013) 1592.

\bibitem{Stoitsov2005} M.V. Stoitsov, J. Dobaczewski, W. Nazarewicz, P. Ring, Comput. Phys. Comm. 167 (2005) 43.

\bibitem{Antonov2017} A.N. Antonov, D.N. Kadrev, M.K. Gaidarov, P. Sarriguren, E. Moya de Guerra, Phys. Rev. C 95 (2017) 024314.

\bibitem{Antonov2018} A.N. Antonov, D.N. Kadrev, M.K. Gaidarov, P. Sarriguren, E. Moya de Guerra, Phys. Rev. C 98 (2018) 054315.

\bibitem{Sarriguren2007} P. Sarriguren, M.K. Gaidarov, E.M. de Guerra, A.N. Antonov, Phys. Rev. C 76 (2007) 044322.

\bibitem{Diep2003} A.E.L. Dieperink, Y. Dewulf, D. Van Neck, M. Waroquier, V. Rodin, Phys. Rev. C 68 (2003) 064307.

\bibitem{Chen2011} L.-W. Chen, Phys. Rev. C 83 (2011) 044308.

\bibitem{Antonov2016} A.N. Antonov, M.K. Gaidarov, P. Sarriguren, E. Moya de Guerra, Phys. Rev. C 94 (2016) 014319.

\bibitem{Danchev2020} I.C. Danchev, A.N. Antonov, D.N. Kadrev, M.K. Gaidarov, P. Sarriguren, E. Moya de
Guerra, Phys. Rev. C 101 (2020) 064315.

\bibitem{Grif57} J.J. Griffin, J.A. Wheeler, Phys. Rev. 108 (1957) 311.

\bibitem{Lalazissis98} G.A. Lalazissis, A.R. Farhan, M.M. Sharma, Nucl. Phys. A 628 (1998) 221.

\bibitem{Geng2004} L.S. Geng, H. Toki, A. Ozawa, J. Meng, Nucl. Phys. A 730 (2004) 80.

\bibitem{Lalazissis2004} G.A. Lalazissis, D. Vretenar, P. Ring, Eur. Phys. J. A 22 (2004) 37 (2004)

\bibitem{ElAdri2019} M. El Adri, M. Oulne, Eur. Phys. J. Plus 135 (2020) 268.

\bibitem{Tagami2020} S. Tagami, J. Matsui, M. Takechi, M. Yahiro, arXiv:nucl-th/2005.13197.

\bibitem{Miller2019} A.J. Miller {\it et al.}, Nat. Phys. 15 (2019) 432.

\bibitem{Ruiz2016} R.F. Garcia Ruiz {\it et al.} Nat. Phys. 12 (2016) 594.

\bibitem{Ozawa2000} A. Ozawa, T. Kobayashi, T. Suzuki, K. Yoshida, I. Tanihata, Phys. Rev. Lett. 84 (2000) 5493.

\bibitem{Warburton90} E.K. Warburton {\it et al.}, Phys. Rev. C 41 (1990) 1447.

\bibitem{Angeli2015} I. Angeli, K. Marinova, J. Phys. G 42 (2015) 055108.

\bibitem{Mutschler2017} A. Mutschler {\it et al.}, Nat. Phys. 13 (2017) 152.

\bibitem{Doornenbal2007} P. Doornenbal {\it et al.}, Phys. Lett. B 647 (2007) 237.

\bibitem{Wang2015} N. Wang, M. Liu, H. Jiang, J.L. Tian, Y.M. Zhao, Phys. Rev. C 91 (2015) 044308.

\bibitem{Antonov2005} A.N. Antonov, D.N. Kadrev, M.K. Gaidarov, E. Moya de
Guerra, P. Sarriguren, J.M. Udias, V.K. Lukyanov, E.V. Zemlyanaya,
G.Z. Krumova, Phys. Rev. C 72 (2005) 044307.

\bibitem{RocaMaza2013} X. Roca-Maza, M. Centelles, F. Salvat, X. Vi\~{n}as, Phys. Rev. C 87 (2013) 014304.

\bibitem{Meucci2014} A. Meucci, M. Vorabbi, C. Giusti, F.D. Pacati, Phys. Rev. C 89 (2014) 034604.

\end{thebibliography}
\end{document}